%% file: article.tex
\documentclass[12pt]{article}

\usepackage{libertinus} 

\usepackage{amsmath}    
\usepackage{amssymb}    
\usepackage{amsthm}     

\newtheorem*{proposition*}{Proposition}

\DeclareMathOperator*{\argmin}{arg\,min}

\usepackage[left=1.5cm,right=1.5cm,top=2.5cm,bottom=2.5cm]{geometry}
\usepackage{placeins}

\usepackage{fancyhdr}
\usepackage{tocloft}
\usepackage{graphicx}   
\usepackage{subcaption} 
\usepackage{tikz}
\usetikzlibrary{matrix, patterns} 
\usepackage{array}

\usepackage{tabularx}   
\usepackage{booktabs}   
\usepackage{makecell}
\usepackage{multirow}
\newcolumntype{C}[1]{>{\centering\arraybackslash}p{#1}}  

\usepackage{algorithm}
\usepackage{algorithmic}

\usepackage[textwidth=4cm, textsize=footnotesize]{todonotes}
\usepackage{comment}
\usepackage{setspace}   
\usepackage{soul,color} 
\usepackage{cancel}     

\usepackage{tasks}      
\usepackage[colorlinks=true, linkcolor=blue, citecolor=blue, urlcolor=blue]{hyperref}
\usepackage{hyperref}



\begin{document}
\bibliographystyle{unsrt}

\begin{center}
{\Large
	{\sc Integrating Structure and Attributes for Transportation Network Partitioning via Optimal Transport}
}
\bigskip

Ketsia Guichard-Sustowski  $^{1}$ \& Loïc Le Marrec $^{2}$ \& Véronique Thelen $^{3}$
\bigskip

{\it
$^{1}$ Univ Rennes, IRMAR - UMR CNRS 6625, F-35000 Rennes, France, ketsia.guichard@univ-rennes.fr\\
$^{2}$ Univ Rennes, IRMAR - UMR CNRS 6625, F-35000 Rennes, France, loic.lemarrec@univ-rennes.fr\\
$^{3}$ Univ Rennes, CREM - UMR CNRS 6211, F-35000 Rennes, France, veronique.thelen@univ-rennes.fr

}
\end{center}
\bigskip


{\bf Abstract.}
Transportation network partitioning is essential for applications such as traffic analysis, simulation, and mobility pattern identification. However, transportation networks combine structural information with heterogeneous operational attributes, ranging from scalar indicators to temporal profiles. Existing approaches generally rely on predefined formulations to integrate these sources of information, limiting the ability to control their relative influence.
This paper proposes a flexible framework for partitioning heterogeneous transportation networks represented as attributed graphs. The proposed methodology relies on a distance-based graph representation and an optimal transport formulation based on the semi-relaxed Fused Gromov--Wasserstein discrepancy, enabling joint consideration of network structure and attributes with explicit control over their trade-off. The proposed methodology is evaluated on two transportation systems with distinct characteristics: an urban road network for traffic-oriented partitioning and a bicycle-sharing system for identifying usage-based communities. Results demonstrate the ability of the framework to adapt the resulting partitions according to different structural and attribute preferences. 

{\bf Keywords}: transportation networks, network partitioning, network representation, multidimensional transportation data, optimal transport, clustering.

\tableofcontents 

\input{sections/1_introduction}
\input{sections/2_litterature_review}
\input{sections/3_representation}
\input{sections/4_proposed_methods}
\input{sections/5_case_studies}

\paragraph{Code Availability} Code to reproduce the results and to apply the proposed methods is available in the following public repository: \href{https://github.com/KetsiaGuichard/ot-transport-network-partitioning/}{Github repository}. The implementation is Python-based and relies on the POT toolbox \cite{flamary2021pot}.

\paragraph{Data Availability} The Origin--Destination trip data for the Santander Cycles public bike-sharing system were provided by Transport for London (TfL). The raw data were obtained from the TfL Open Data service and used in accordance with the applicable data licence. The processed traffic datasets of Châteaubourg used in this study are publicly available in this \href{https://doi.org/10.5281/zenodo.21602917}{Zenodo repository}.

\paragraph{Acknowledgement} The authors gratefully acknowledge Ioana Gavra for the insightful discussions on data representations and for the continuous support provided throughout this work. The authors also thank the association Agis Ta Terre for its keen interest and active involvement in the field, as well as Neovya for its valuable support and contribution to this study.

\paragraph{Fundings} This work was conducted within the France 2030 program, Centre Henri Lebesgue ANR-11-
LABX-0020-01. This project was supported by the participatory research platform of TISSAGE - Science avec et pour la société.

\bibliography{references}

\end{document}

%% file: sections/1_introduction.tex
\section{Introduction}

Transportation infrastructures may be represented as networks, making network analysis an important research topic in transportation science. In particular, network approaches have been widely used to characterize transportation systems and investigate their structural properties \cite{lin2013complex}. A central issue in this context is the choice of an appropriate network representation. For example, road networks can be represented through a primal representation, where road segments are links, or through a dual representation, where roads are modeled as nodes \cite{porta2006anetwork,porta2006bnetwork}. Similar considerations have been explored for public transportation networks, showing that network representations should be adapted to the objective of the analysis \cite{von2007network}.

Beyond the characterization of global properties such as centrality measures \cite{sun2018role}, clustering coefficients, and degree distributions \cite{sienkiewicz2005statistical}, network analysis also aims at identifying meaningful substructures within transportation systems. Network partitioning addresses this challenge by dividing a network into groups of nodes or regions sharing common characteristics. Such partitions support several transportation applications, including distributed traffic simulation \cite{leclercq2021enforcing}, road network partitioning \cite{ji2012spatial}, and community detection in mobility systems \cite{yildirimoglu2018identification}.

The increasing availability of heterogeneous transportation data has broadened the notion of similarity in network partitioning. Transportation networks are no longer characterized solely by their topology, but also by multiple descriptors collected from different sources. Road networks, for instance, combine structural information describing connectivity and spatial organization with operational attributes that may take various forms, ranging from scalar indicators to temporal traffic profiles \cite{crawford2017statistical} or travel time distributions \cite{fosgerau2012valuing}. Consequently, transportation networks can be represented as attributed graphs, where both structural relationships and heterogeneous attributes contribute to the characterization of networks. This diversity of information motivates the development of partitioning frameworks capable of integrating different types of attributes while controlling their relative influence.

Existing approaches have demonstrated the relevance of combining structural and operational information for transportation network partitioning. For instance, \cite{saeedmanesh2016clustering} proposed the Snake similarity measure, which incorporates directional flow information while considering the spatial organization of transportation networks. More generally, attributed network partitioning approaches rely on different formulations to combine structural and attribute-based similarities. However, the relative importance assigned to these two sources of information is often implicitly determined by the selected representation or similarity formulation. Since transportation applications may pursue different objectives, a flexible framework allowing explicit control over the trade-off between structural and attribute information remains desirable.

This paper proposes a new flexible framework for partitioning attributed transportation graphs, based on optimal transport. The proposed methodology relies on the semi-relaxed Fused Gromov--Wasserstein discrepancy, which enables the comparison of graphs by jointly considering their internal structure and node attributes. By controlling the contribution of structural and attribute information, the proposed framework provides a flexible and versatile approach, adaptable to different transportation partitioning objectives.

Throughout this paper, an urban road network is used as a running example to illustrate the construction of attributed transportation graphs and the impact of representation choices on the resulting partitions.

\newpage
The main contributions of this work are:
\begin{enumerate}
    \item A general framework for representing heterogeneous transportation networks as attributed graphs is proposed, leveraging a distance-based formulation to integrate structural information with heterogeneous transportation attributes. This formulation enables the combination of attributes with different natures, including complex descriptors such as temporal profiles, distributions, and scalar features, within a unified representation;
    \item An optimal transport-based partitioning methodology is introduced, relying on the semi-relaxed Fused Gromov--Wasserstein discrepancy, which provides explicit control over the trade-off between structural and attribute information;
    \item The applicability and flexibility of the proposed framework are demonstrated on two transportation systems with distinct partitioning objectives: urban road network partitioning for traffic analysis and community detection in bicycle-sharing systems.
\end{enumerate}

The remainder of this paper is organized as follows. Section 2 introduces the main motivations and applications of transportation network partitioning and reviews existing approaches. Section 3 discusses the representation of transportation networks as attributed graphs, including the construction of distance-based representations and their implications for partitioning. Section 4 presents the proposed optimal transport-based methodology and its comparison with existing approaches. Section 5 evaluates the flexibility of the framework through two case studies involving an urban road network and a bicycle-sharing system.

\paragraph{Problem Description and Notations}

A graph $G = (V, E)$ consists of $n$ nodes $V = \{v_1, \dots, v_n\}$ and $m$ edges $E = \{e_{ij}\} \subseteq V \times V$. The topological structure of $G$ can be defined by the $n\times n$ adjacency matrix $A = [a_{ij}]_{n \times n}$, with $a_{ij} = 1$ if there is an edge between $v_i$ and $v_j$ ($e_{ij} \in E$) or 0 otherwise. Graphs can be directed ($e_{ij} \neq e_{ji}$) or undirected ($e_{ij} = e_{ji}$). Both vertices and edges can have weights associated with them, represented by a weight matrix $W = [w_{ij}]_{n \times n}$. Although a soft clustering problem could be considered, the present work focuses on a hard clustering setting. In this case, the clustering of an attributed graph aims to partition the set of nodes $V$ into $k$ subsets $C_1, \dots, C_k$ of $G$ such that $\cup^k_{i = 1} C_i = V$ and $C_i \cap C_j = \emptyset$ for $i \neq j$.

%% file: sections/2_litterature_review.tex
\section{Role of Network Partitioning in Transportation Systems}

As mentioned in the introduction, transportation network partitioning addresses a variety of objectives, and the relevant criteria strongly depend on the considered application. For parallelization-related applications, for instance, a key requirement is often to obtain a limited number of partitions with comparable computational loads. Conversely, in community detection applications, partition sizes may naturally vary, highlighting that criteria suitable for one objective may be irrelevant or even conflicting for another. The following section reviews these different application contexts and provides an overview of the main partitioning approaches employed for these purposes.

\subsection{Network Partitioning Applications}

\paragraph{Parallelization for Large Networks} In large partitioning networks, analysis, optimization or simulation tasks can be challenged by the size of the network, leading to the need for dividing the traffic network into smaller subnetworks. For instance, road traffic simulation, especially at a microscopic or nanoscopic scale, may require distributed or parallel computing environments, whose performance depends on the spatial decomposition of the network \cite{potuzak2022current}. This motivation for parallelization directly shapes the optimization problem, for instance by aiming to minimize the number of border roads or the number of neighboring subnetworks. Similarly, decomposition strategies have also been investigated for static traffic assignment problems \cite{jafari2017decomposition}.

In this case, partitioning can be performed by dividing roads based on spatial information, for example by splitting the map into grids or regions of similar size \cite{berger1987partitioning}. The main drawback of such approaches is that they can generate excessive communication between zones, thereby reducing parallel efficiency. To address this issue, alternative methods treat the road network as a graph and apply graph-partitioning techniques that consider either only network information \cite{ventresque2012spartsim} or a combination of network topology and vehicle-specific information \cite{liu2024parallel}.


\paragraph{Road Network Partition} In the context of traffic management, partitioning concepts are also employed, with the objective of identifying distinctly congested areas in order to design region-specific traffic control strategies \cite{anwar2014spatial}. \cite{ji2012spatial} summarizes this objective for road network partitioning as minimizing the variance of link densities within each cluster, subject to constraints on a limited number of clusters and their spatial compactness. The quality of a network partition can be evaluated using different criteria, such as its consistency with theoretical concepts like the Macroscopic Fundamental Diagram \cite{lin2020road}. Studies on aggregated traffic modeling have highlighted that urban networks exhibit significant spatial heterogeneity, motivating the identification of homogeneous regions that can support hierarchical traffic control strategies while preserving relevant macroscopic traffic characteristics \cite{ramezani2015dynamics}. Road network partitioning can be performed using different types of traffic information, typically including density and/or speed, or even data from multiple modes of transportation \cite{johari2023mode}.

\paragraph{Community Detection} Beyond large-scale network optimization, transportation network partitioning is also employed for analytical purposes, such as understanding usage patterns or identifying zones with similar behavior. This type of analysis is often referred to as community detection, although the broader objective may be to partition the entire network into meaningful subnetworks. In this use case, when the focus is on usage patterns, spatial interactions may be less important than similarities in mobility behavior, which has implications for network construction. For instance, \cite{liu2024exploring} define edge weights based on a traffic similarity measure rather than geographical distance. Other community detection approaches, however, rely on networks in which edges are defined by the intensity of flows between origin and destination zones \cite{huang2018comparing}. As in the case of parallelization, this type of analysis can also be applied to multiple modes of transportation \cite{xie2024analyzing}.

Community detection has been widely applied in bicycle-related analyses, such as bicycle network design \cite{akbarzadeh2018designing}, and even more extensively in the study of bicycle-sharing systems. Clustering stations or zones can be useful for identifying localized usage patterns \cite{zaltz2013structure}, better understanding the socio-demographic factors influencing service accessibility \cite{lee2025exploring}, and distinguishing different user communities according to peak and off-peak periods \cite{sun2023study}.

\subsection{Existing Approaches for Transportation Network Partitioning} \label{method_state_of_art}

Table \ref{table:clustering_methods} presents a classification of partitioning methods that are commonly used for different applications in transportation networks. Note that the clustering families in this classification may be related or may exhibit overlaps. For instance, spectral clustering and $k$-means-based clustering, particularly kernel $k$-means, share the same mathematical foundation \cite{filippone2008survey}. The proposed clustering families are presented below to provide insights into their behavior.

\begin{table}[ht]
\begin{tabular}{|C{3cm}|C{6.3cm}|C{4.5cm}|C{3.2cm}|}
\hline
Clustering family & Method & Transportation references & Primary application \\
\hline
cut-based & Kernighan-Lin \cite{kernighan1970efficient}, METIS \cite{karypis1997metis} & \cite{ahmed2016partitioning} \cite{ventresque2012spartsim} \cite{xu2017graph} & Parallelization \\
\hline
\multirow{3}{*}{spectral} & N-cut \cite{shi2000normalized} and variants & \cite{anwar2014spatial}  \cite{bell2017investigating} \cite{ji2012spatial} \cite{liu2024parallel} & \multirow{3}{*}{\makecell[c]{Road Network\\Partitioning}} \\
    \cline{2-3} & evolutionary spectral clustering \cite{chi2007evolutionary} &  \cite{al2022partitioning} & \\
    \cline{2-3} & SymNMF \cite{kuang2015symnmf} & \cite{liu2019new} \cite{saeedmanesh2016clustering} & \\
\hline
\multirow{4}{*}{modularity-based} & Fast Greedy \cite{clauset2004finding} & \cite{liu2024exploring} \cite{zaltz2013structure}& \multirow{4}{*}{\makecell[c]{Community\\Detection}} \\
    \cline{2-3} & Louvain \cite{blondel2008fast} & \cite{akbarzadeh2018designing} \cite{borgnat2011shared} \cite{gurtner2014multi}
    \cite{huang2018comparing}
    \cite{lee2025exploring} \cite{liu2024exploring} & \\
    \cline{2-3} & COMBO \cite{sobolevsky2014general} & \cite{huang2018comparing} \cite{li2016identifying} & \\
    \cline{2-3} & Leiden \cite{traag2019louvain} & \cite{kim2025detecting} \cite{liu2024exploring}& \\
\hline
\multirow{2}{*}{\makecell[c]{random walk}} & Infomap \cite{rosvall2008maps} & \cite{gurtner2014multi} \cite{huang2018comparing} & \multirow{2}{*}{\makecell[c]{Community\\Detection}}\\
    \cline{2-3} & Walktrap \cite{pons2006computing} & \cite{huang2018comparing} \cite{sun2023study} & \\
\hline
hierarchical & explicit hierarchical method \cite{simon1997good} & \cite{wei2010improved} \cite{xie2024analyzing} & Various applications\\
\hline
$k$-means-based & $k$-means \cite{hartigan1979algorithm} and variants \cite{okabe2008generalized} & \cite{lin2020road} \cite{liu2021spatio} \cite{yin2020victs} & Various applications \\
\hline
\end{tabular}
\caption{Classification of graph partitioning methods for transportation networks.}
\label{table:clustering_methods}
\end{table}

\paragraph{Cut-based Methods} In this work, we refer to cut-based methods as approaches that aim to find a partition of the graph in which the weights of edges between different groups are minimized. This optimization problem also appears in spectral clustering, although in a relaxed way, and in hierarchical methods, which produce an explicit final hierarchy without node-moving or refinement procedures. A commonly used method is the Kernighan-Lin algorithm \cite{kernighan1970efficient}, which starts with an initial partition and then iteratively swaps pairs of nodes between partitions to reduce the cut cost, defined based on the weights of edges between partitions. However, their heuristic nature and tendency to produce hard, balanced partitions can limit their effectiveness on complex networks with overlapping or hierarchical structures.

\paragraph{Spectral Methods} Spectral clustering \cite{ding2024survey} relies on the eigenvectors and eigenvalues (the "spectrum") of the graph Laplacian matrix. Partitioning a graph into $k$ components amounts to selecting the $k$ eigenvectors associated with the $k$ smallest eigenvalues, yielding a low-dimensional embedding. In this subspace, the clusters are geometrically easier to separate and can be identified using a traditional clustering method, such as $k$-means. Since working with unnormalized Laplacian can lead to unbalanced partitioning, spectral methods often employ a normalized version of the Laplacian (such as $I - D^{-1/2}WD^{-1/2}$), which helps to balance the resulting clusters. The performance of this family of method strongly depends on the choice of similarity representation, embedding dimensionality, and regularization parameters, which may limit their robustness across diverse graph structures.

\paragraph{Modularity-based Methods} The concept of modularity was introduced by \cite{newman2004finding} and compares the actual number of edges within a community to the expected number of edges if edges were distributed randomly while preserving the same degree distribution. Modularity can be either positive or negative, with positive values indicating a potential community structure. Based on the maximisation of modularity, various algorithms have been proposed to partition graphs. The Louvain method (also known as Fast Unfolding \cite{blondel2008fast}) is one of the most widely used approaches, alternating between local node movements and multilevel aggregation, and offering higher scalability. However, modularity-based methods are limited by their focus on graph topology, which can lead to communities that are well connected structurally but poorly aligned with the underlying attribute-based organization.

\paragraph{Random Walk Methods} A random walk on a network is a stochastic process in which a walker moves from one node to another according to transition probabilities that depend on edge weights. Through its exploration of the graph, the walker tends to spend longer periods in densely connected regions, thereby reflecting the presence of communities. Their performance can be sensitive to graph properties such as degree distribution and walk parameters, limiting their robustness on large-scale, heterogeneous, or attributed networks.

\paragraph{Hierarchical Methods} Hierarchical methods are defined as approaches that iteratively merge (agglomerative methods) or split (divisive methods) nodes based on a measure of similarity or distance. An example is the recursive bisection of graphs \cite{simon1997good}, which iteratively splits a graph into two subgraphs at each step while minimizing the number of edge cuts. By construction, hierarchical methods can capture multi-scale structures in networks, and some variants have even been designed for multilayer networks, which is often the case in transportation research, for example when identifying overlapping communities \cite{liu2018finding}. Hierarchical methods could be limited limited by their recursive clustering process, which can propagate early decision errors across the hierarchy.

\paragraph{$k$-means-based Methods} The $k$-means method \cite{hartigan1979algorithm} aims to partition data into a fixed number of clusters by iteratively assigning each data point to the nearest cluster centroid and updating the centroids so as to minimize within-cluster variance. While the traditional implementation (Lloyd algorithm \cite{lloyd1982least}) is not directly applicable to graphs, where distances are constrained by the network structure, several alternatives have been proposed, like Network Voronoi Diagrams \cite{okabe2008generalized}. Another approach consists in selecting a representation of the original data that does not depend on the network structure, allowing $k$-means to be applied. However, this latest approach is limited by its reliance on vector-space representations, which may not adequately capture the complex topology of graphs.

Most existing approaches for transportation network partitioning are designed for graphs without attributes and mainly rely on topological information. Although they can be extended to attributed graphs, their performance strongly depends on the chosen representation of node attributes. A generic partitioning framework, capable of handling heterogeneous attributed graphs while remaining easily adaptable to different application settings, is therefore needed. The method introduced in Section~\ref{srfgw} is designed with this objective in mind. However, such a method requires beforehand a flexible representation framework capable of handling a broad range of attribute types, which is discussed in the following section.

%% file: sections/3_representation.tex
\section{Representing Transportation Networks as Attributed Graphs} \label{representation}

While transportation network partitioning has been applied to a broad range of problems, an important methodological question concerns how transportation systems are represented prior to partitioning. In practice, the partitioning outcome strongly depends on the way topological structures, spatial relationships, and transportation attributes are encoded within the graph. A typical workflow first consists in constructing a graph representation of the transportation system, potentially enriched with additional information such as traffic flows, speeds, or temporal usage patterns. Similarity or distance measures are then defined in order to capture relationships between network elements, before applying a partitioning algorithm.

\subsection{Running Example: Road Network Dataset}\label{running_example}

To illustrate the representation choices discussed in the following sections, a real-world urban road network enriched with Floating Car Data (FCD) is introduced as an illustrative case study. At this stage, the objective is not to analyze partitioning results themselves, but rather to provide a concrete support for discussing graph construction, similarity modeling, and attributed transportation graph representations. The interpretation of the resulting partitions and the methodological evaluation of the proposed framework are deferred to the case study section (section \ref{case:roadnetwork}).

\begin{minipage}{0.6\textwidth}
The city used in this reference example is Châteaubourg (Brittany, France), a small town of approximatively 7,500 inhabitants that experiences recurrent congestion issues. Despite its size, the city has a significant employment base and lies between two major urban centers in the region. This city was selected due to an ongoing citizen project, that aims to measure traffic to better understand car mobility in low-density areas and
support informed public policy decisions.

\smallskip
The road network of Châteaubourg (Figure \ref{ctb_network})
is represented in our framework by a graph composed of 832 streets segments (corresponding to portions of streets), connected through 748 intersections. These intersections correspond not only to physical junctions, such as roundabouts or crossroads, but also to simple connections between consecutive streets portions. Specifically, each street is divided into subsections between intersections connecting to that street, in order to ensure a graph-based representation. As a result, the lengths of road segments can vary substantially, ranging from 7 meters in the city center to 5 kilometers in rural areas.
\end{minipage}
\hfill
\begin{minipage}{0.35\textwidth}
    \centering
    \centering
    \includegraphics[width=0.8\linewidth]{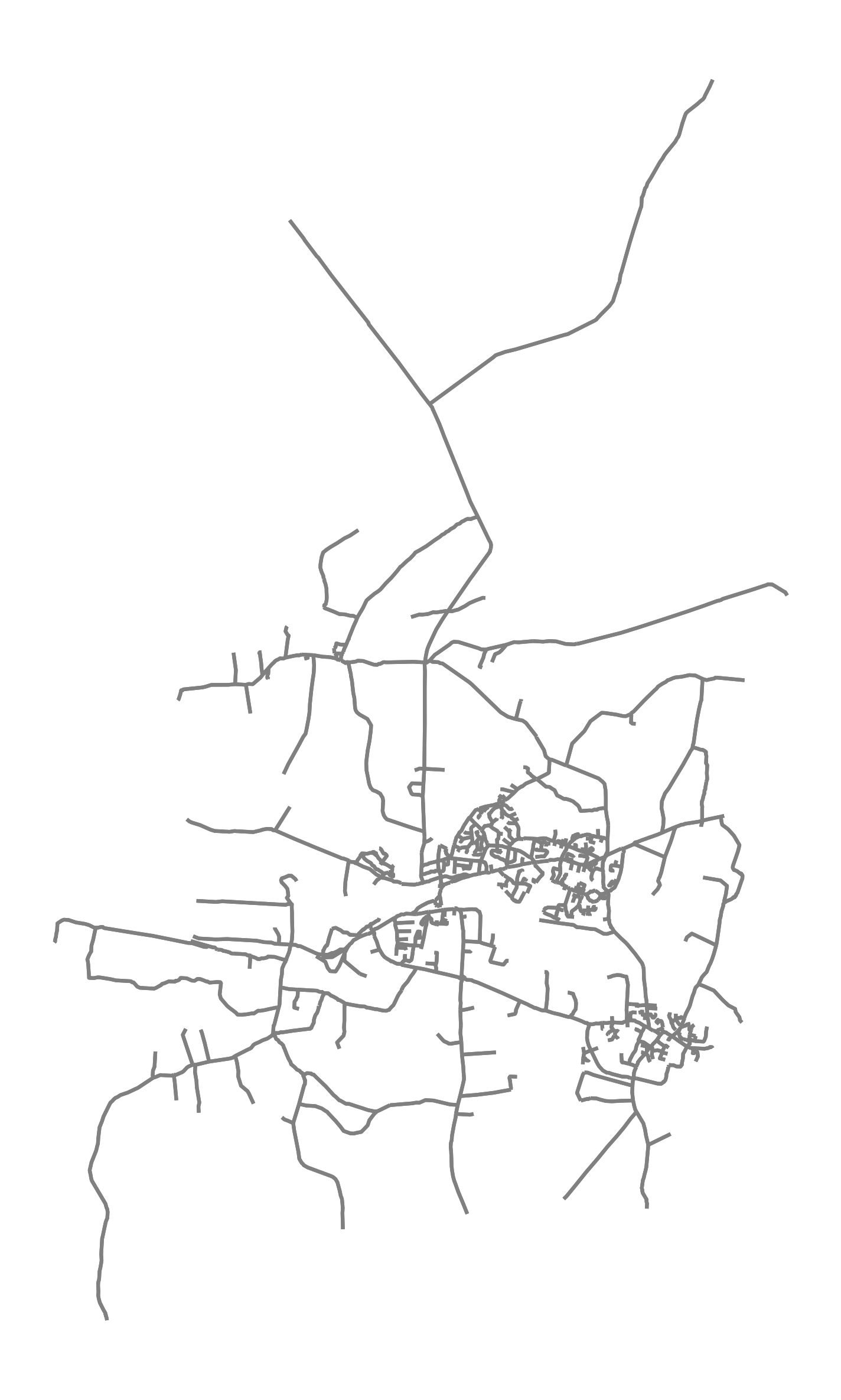}
    \captionof{figure}{Road network of Châteaubourg}
    \label{ctb_network}
\end{minipage}

Raw FCD data collected in 2022 provide speed statistics and vehicle counts. In this study, a pre-processed and resampled dataset is used, containing traffic curves (number of vehicles per 15-minute interval) and speed histograms (with 5 km/h bins), aggregated by road segment and direction. These attributes provide richer information than simple summary statistics: traffic curves capture temporal patterns such as commuting behavior, while speed histogram shapes can reveal specific regimes, such as congestion effects (right-skewed distributions) or distinct speed regimes (bimodal distributions).

Speed histograms are aggregated over the entire observation period to obtain robust representations. Traffic curves are aggregated into a representative weekly profile at a 15-minute resolution. Using an average week rather than a specific day limits sampling bias while capturing recurrent temporal patterns, such as reduced traffic during weekends.

\smallskip
\begin{minipage}{0.3\textwidth}
Traffic counts are normalized using min-max scaling for each road section and direction to prevent the analysis from only reflecting overall traffic volume differences and to enable a more fine-grained interpretation. Finally, although data are available for both directions of travel, we adopt a modeling choice in which each road segment is represented by a single node. The associated attributes combine information from both directions, while preserving their directional distinction. This choice is motivated by the ultimate objective of partitioning streets, while also simplifying the graph construction. Figure \ref{ctb_attributes_illus} provides an illustration of the traffic data for a given street.
\end{minipage}
\hfill
\begin{minipage}{0.65\textwidth}
    \centering
    \centering
    \includegraphics[width=0.90\linewidth]{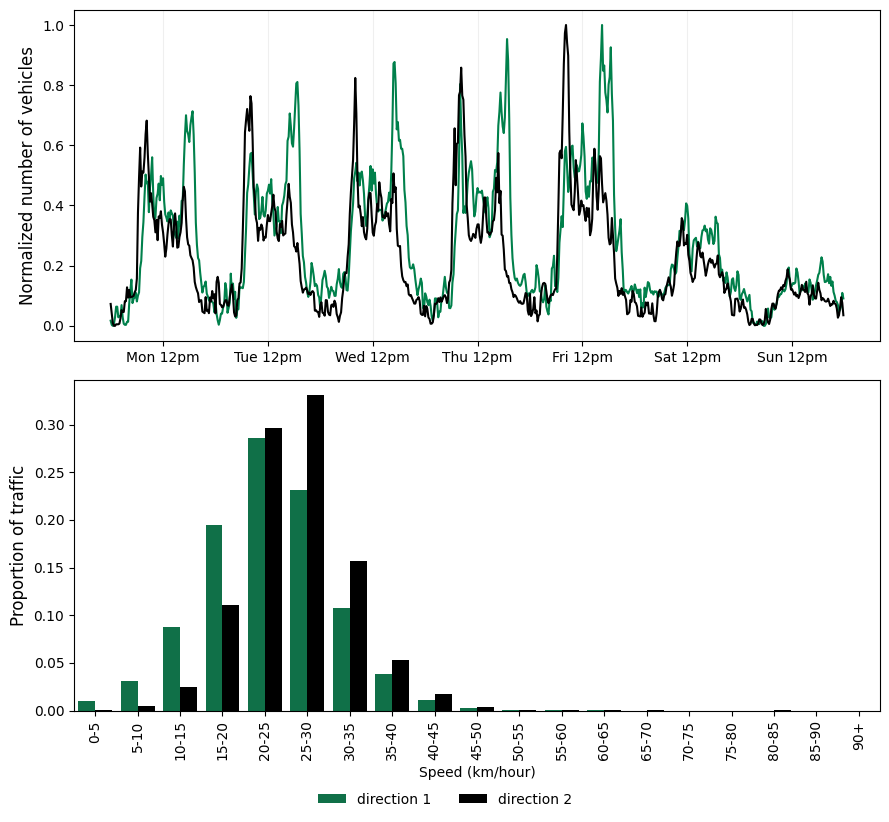}
    \captionof{figure}{Illustration of the attributes associated with a road segment: normalized traffic and speed from both directions}
    \label{ctb_attributes_illus}
\end{minipage}

In the remainder of the methodological sections, this use case is used to illustrate the different representations and methods considered. The interpretation of the results is left to the dedicated use case section (Section~\ref{case:roadnetwork}).

\subsection{Graph Construction}

For transportation partitioning, the graph can be derived from a spatial network, such as a road or transportation network. As highlighted in the introduction, even in this case, the representation should be carefully chosen depending on the research objective, using either a primal or dual representation. Alternatively, the graph can be constructed between spatial points, such as stations or zones, where edges represent origin-destination flows.  In both cases, the graph can be either directed or undirected, with edges representing one-way or two-way roads or links.

In our example, intersections serve as nodes and street segments as edges. However, since the goal is to partition the streets rather than the intersections, the dual graph representation is adopted \cite{porta2006bnetwork}. In the dual graph, each road segment is represented as a node, and an undirected edge is established between two nodes whenever their corresponding street segments share at least one intersection in the original network. These edges thus encode the adjacency between road segments. This dual network is composed of 832 nodes (former street segments) and 2166 edges.

As in the primal representation, the dual construction does not preserve node coordinates within a geographic reference system. Moreover, since edges in the dual graph represent intersections between street segments, they do not inherently carry a notion of length. To retain topological distance information, a weight equal to the mean distance of its two endpoints (i.e., the two street segments it connects) is assigned to each edge of the dual graph. A distance matrix can therefore be computed between every pair of nodes in the graph, using shortest-path distances, yielding a structural distance matrix denoted $\mathbf{D_S} = [d_S(v_i, v_j)]_{1 \leq i,j \leq n}$.

\subsection{Similarity and Distance Modeling}

Most transportation partitioning applications presented previously do not rely solely on the network topology, but also incorporate additional information associated with nodes or edges. As a result, constructing an attributed transportation graph generally requires defining a notion of similarity or distance capable of capturing relationships within the attributes.

In the simplest cases, transportation attributes may be represented using scalar quantities such as average traffic flows, densities, or speeds, for which conventional metrics such as the Euclidean distance can be employed. However, transportation systems often exhibit more complex behaviors that cannot be adequately summarized by scalar indicators alone. In the running case study considered throughout this section, traffic information is therefore represented using richer data structures, including traffic curves and speed histograms, in order to better characterize mobility dynamics and traffic variability.

A contribution of this work is the adoption of distances as a common representation of transportation attributes. By representing attributes through pairwise distance matrices, any attribute defined on a metric space can be incorporated using its notion of distance and combined within a unified framework. This choice is motivated by the wide diversity of attribute types encountered in transportation networks and is illustrated below using the traffic curves and speed histograms of the running example.

\paragraph{Traffic Curves} Traffic data may exhibit slight time shift in their peaks across different geographical zones, but the primary objective is to detect similarity in shape. Therefore, Dynamic Time Warping (or DTW - \cite{senin2008dynamic}) is employed, denoted $d_{\text{dtw}}(f_i, f_j)$, where $f_i$ and $f_j$ are the traffic curves associated to nodes $i$ and $j$, respectively, and are treated as discrete time series. Formally, letting $X^{(i)} = (x^{(i)}_1,\dots,x^{(i)}_m)$ and $X^{(j)} = (x^{(j)}_1,\dots,x^{(j)}_m)$ denote the sampled values of $f_i$ and $f_j$, DTW computes an optimal alignment between $X^{(i)}$ and $X^{(j)}$ by allowing non-linear warping along the time axis under constraints. It determines a warping path $\pi$ that minimizes the cumulative distance $\sum_{(t,s)\in \pi} d(x^{(i)}_t, x^{(j)}_s)$. This provides a meaningful similarity measure even when time series are temporally shifted, for instance when traffic peaks occur at slightly different times on two roads while preserving the same overall pattern.

\paragraph{Speed Histograms} Speed histograms require a distance metric suitable for comparing distributions. In this study, histogram data are compared using the Wasserstein-1 distance, denoted $W_1(\mathbf{{h}^{(i)}}, \mathbf{h}^{(j)})$, where $\mathbf{h}^{(i)}$ (resp. $\mathbf{h}^{(j)}$) is the discrete distribution associated to node $i$ (resp. node $j$). 
\begin{equation*}
W_1(\mathbf{h}^{(i)}, \mathbf{h}^{(j)}) 
= \frac{1}{T} \sum_{t=1}^{T} \left| h^{(i)}_t - h^{(j)}_t \right|
\end{equation*}

Wasserstein-1 defines a distance between distributions based on optimal mass transport, taking into account the underlying geometry. Unlike pointwise metrics such as the Euclidean distance, it is sensitive to shifts and captures differences in shape. 

\paragraph{Adaptation for both directions}
Depending on whether a street is one-way or two-way, each node in the graph is characterized by one or two speed histograms and one or two weekly traffic curves. Consequently, the attribute distances used in our simulations must be adapted to compare subsets of attributes. Let $\mathcal{X}$ denote the metric space associated with the considered attribute type, equipped with the distance $d_{\mathcal{X}}$. The Hausdorff distance is specifically designed to measure distance between two subsets $A$ and $B$, and numerous variations have been introduced with the purpose of object matching \cite{dubuisson1994modified}. In this study, the following variant is adopted:
\begin{equation*}
    d(a, B) = \min_{b \in B} d_{\mathcal{X}}(a,b), \quad 
    \textrm{HD}(A, B) = \frac{1}{2} \Big( \max_{a \in A} d(a, B) + \max_{b \in B} d(b, A) \Big)
\end{equation*}

In our setting, $A$ denotes the set of traffic curves (respectively speed histograms) associated with a road segment, and $B$ the corresponding set for another segment. The distance $d_{\mathcal{X}}$ is defined as $d_{\mathrm{dtw}}$ (respectively $W_1$).

This formulation ensures that both traffic directions associated with a road segment are treated symmetrically, while still penalizing discrepancies between the corresponding subsets of attributes. The resulting attribute-based distances between nodes $v_i$ and $v_j$ are denoted by $d_{\textrm{flow}}(v_i, v_j)$ for weekly traffic flow curves and $d_{\textrm{speed}}(v_i, v_j)$ for speed histograms.

\paragraph{Combine Attributes}

Since the variances of the traffic flow and speed-based distances are relatively large, a square-root transformation is applied to each distance matrix in this application. This transformation helps prevent trivial clusterings, such as a separation driven solely by extreme differences between low-speed and high-speed roads, and promotes a more balanced contribution of the different types of attributes. To enable comparability across distance types, the transformed attribute distances are further rescaled using min–max normalization, yielding $\tilde{d}_{\mathrm{flow}}(v_i, v_j)$ and $\tilde{d}_{\mathrm{speed}}(v_i, v_j)$, respectively.

It is worth noting that the same square-root transformation is applied to the shortest-path distance matrix computed on the dual road network, as this network exhibits large variability in edge lengths, as discussed previously.

Once these transformations are applied, the different attribute distances can be combined at the road level:
\begin{equation*}
d_A(v_i, v_j) = \beta \cdot \tilde{d}_{\textrm{flow}}(v_i, v_j) + (1 - \beta) \cdot \tilde{d}_{\textrm{speed}}(v_i, v_j).
\end{equation*}

$\beta$ is fixed to $\frac{1}{2}$ in the running example, but this weighting strategy can be adapted depending on the application.

\subsection{Combining Attributes and Structure}

After constructing the graph and defining appropriate distance measures for the attributes, the next step consists in specifying how topological and attribute information are combined within a unified representation prior to the application of a partitioning method.

For instance, \cite{liu2024parallel} propose a methodology to incorporate vehicle-specific information into the partitioning of the road network by including capacity, topological information and exchange volume in the weight function. This weighting adjacency strategy is quite common, sometimes combining topological and attribute information (as in the previous example), but more often considering only attribute information in the weights, especially when the network is build through an origin-destination matrix (for instance \cite{akbarzadeh2018designing}). In both cases, the attribute information is generally limited to a single attribute type, whether scalar (e.g., average speed) or functional (e.g., traffic profiles), rather than combining heterogeneous attributes. It should be noted that, although uncommon, some studies consider more complex forms of attributes, such as \cite{liu2024exploring}, which use DTW to incorporate traffic curve information as a weighting parameter in the graph.

A specific representation commonly used for road network partitioning deserves particular attention due to its original combination of topological and traffic information, its widespread adoption in the literature, and its explicit consideration of traffic attributes. Snake \cite{saeedmanesh2016clustering} constructs a representation by growing "Snakes", sequences of connected links, by iteratively adding adjacent links with similar traffic properties, minimizing the variance of the attributes. A similarity matrix is then computed from the resulting sequences, forming the basis for a clustering algorithm. In the original paper, this similarity matrix is subsequently used as input to a spectral clustering method. Several parameters act as weighting factors between topological and structural information, notably the Snake size (shorter Snakes constrain the clustering to more local structures) and a weighting coefficient associated with the spatial component, which promotes more compact clusters. 

\subsection{Implications for Network Partitioning}

While the presented representations differ in their philosophy and formulation, their main impact is expected to concern the resulting partition, independently of the partitioning method used. Therefore, the different representation choices presented above are illustrated through the running case study: weighted adjacency, combined distance/similarity matrices, and Snake similarity with different Snake lengths. The resulting partitions obtained with a similar clustering method using these different representations are then compared.

The weighted adjacency strategy consists in considering the graph while replacing the purely structural edge length by a combination of structural and attribute-based distances, weighted by a parameter $\alpha \in [0,1]$, which controls the relative contribution of the structural component. A weighted adjacency matrix, denoted $\mathbf{A}_{\alpha}$, is then defined through its entries as follows:
\begin{equation*}
a_{ij} =
\begin{cases}
\alpha \, d_{\text{S}}(v_i, v_j) + (1 - \alpha)\, d_{\text{A}}(v_i, v_j) & \text{if there exists an edge between nodes } i \text{ and } j, \\
0 & \text{otherwise.}
\end{cases}
\end{equation*}

The combined distance matrix follows the same idea, but extends this combination beyond adjacent nodes to all pairs of nodes, yielding a new pairwise distance matrix $\mathbf{D}_{\alpha}$, where each entry is defined as $d_{\alpha}(v_i, v_j) = \alpha \, d_{\text{S}}(v_i, v_j) + (1 - \alpha)\, d_{\text{A}}(v_i, v_j)$. Since some partitioning methods, like spectral clustering, don't operate directly on a distance matrix, a similarity matrix $\mathbf{S}_{\alpha} = [s_{ij}]_{1 \leq i,j \leq n}$ can instead be derived from $\mathbf{D}_{\alpha}$ using a Gaussian kernel, where similarity decreases exponentially with the squared distance. The kernel scale is controlled by a bandwidth parameter $\sigma$, typically estimated from the data (here taken as the median of pairwise distances). 
\begin{equation*}
s_{ij} = \exp\left(-\frac{d_{\alpha}(v_i, v_j)^2}{2\sigma^2}\right).
\end{equation*}

The Snake similarity strongly depends on either the Snake length or a weighting coefficient, both of which control the balance between topology and attributes by determining the spatial scale at which the network is analyzed. For simplicity, only the Snake length parameter is considered here.

\paragraph{Choice of $\alpha$}
Both the weighted adjacency and the combined distance/similarity representations depend on the parameter $\alpha$, which controls the relative weight assigned to the structural component. This parameter can be chosen according to the objective of the analysis and the desired balance between structural and attribute information.

\begin{figure}
    \centering
    \includegraphics[width=\textwidth]{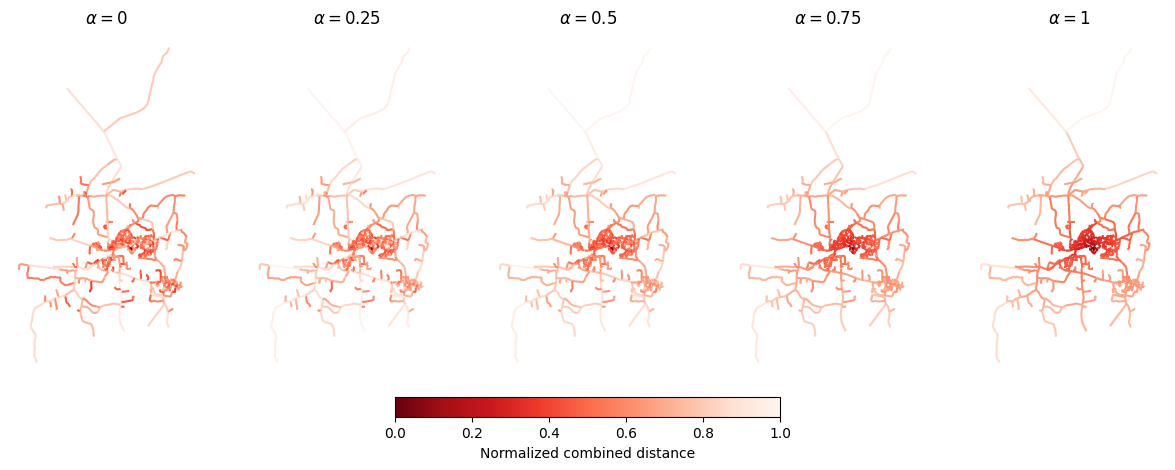}
    \caption{Normalized combined $\alpha$-distance to a road located in the city center for different values of $\alpha$.}
    \label{alpha_illus}
\end{figure}

Figure~\ref{alpha_illus} illustrates the effect of different values of $\alpha$ on the combined distance for a road located in the city center: the lighter the red, the larger the distance to the reference road. One may observe that low values of $\alpha$ lead to a distance mainly driven by street categories, with suburban roads appearing very distant from the city-center reference. Conversely, for large values of $\alpha$, the distance naturally increases as one moves farther away from the city center.

In addition to choices guided by the application, this parameter can also be selected to optimize a partitioning criterion. For the remainder of this section, $\alpha$ is set to $0.5$.

\begin{figure}
    \centering
    \includegraphics[width=\textwidth]{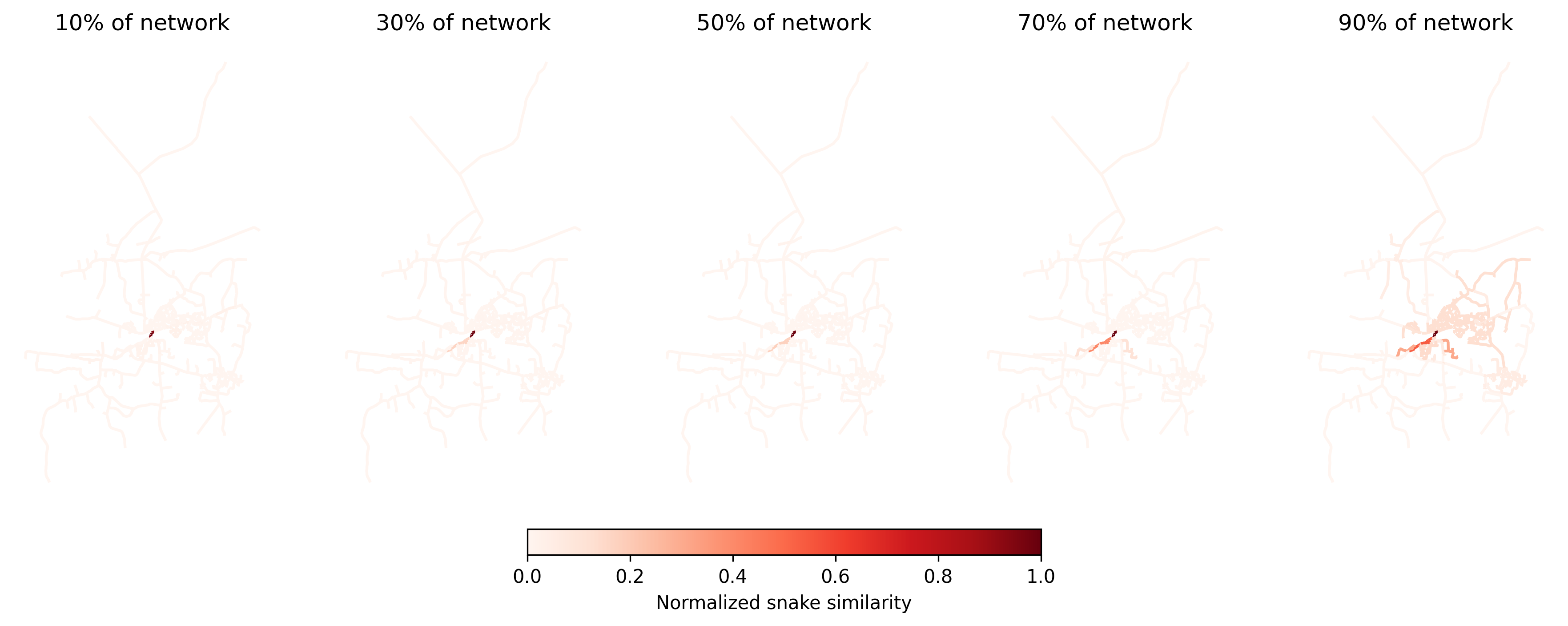}
    \caption{Normalized Snake similarity to a road located in the city center for different Snake lengths, expressed as a proportion of network length.}
    \label{Snake_illus}
\end{figure}

\paragraph{Choice of Snake length}
Figure~\ref{Snake_illus} shows how the Snake length influences the similarity distribution: short Snakes lead to identifying only very nearby roads as similar, while the rest of the network is considered dissimilar. With very long Snakes (covering almost the entire network), this binary effect is smoothed, but the compactness of the set of similar roads remains relatively strong, and the attributes appear to play a smaller role than in the previous combined representations.

\paragraph{Impact on clustering}

\begin{figure}
    \centering
    \includegraphics[width=\textwidth]{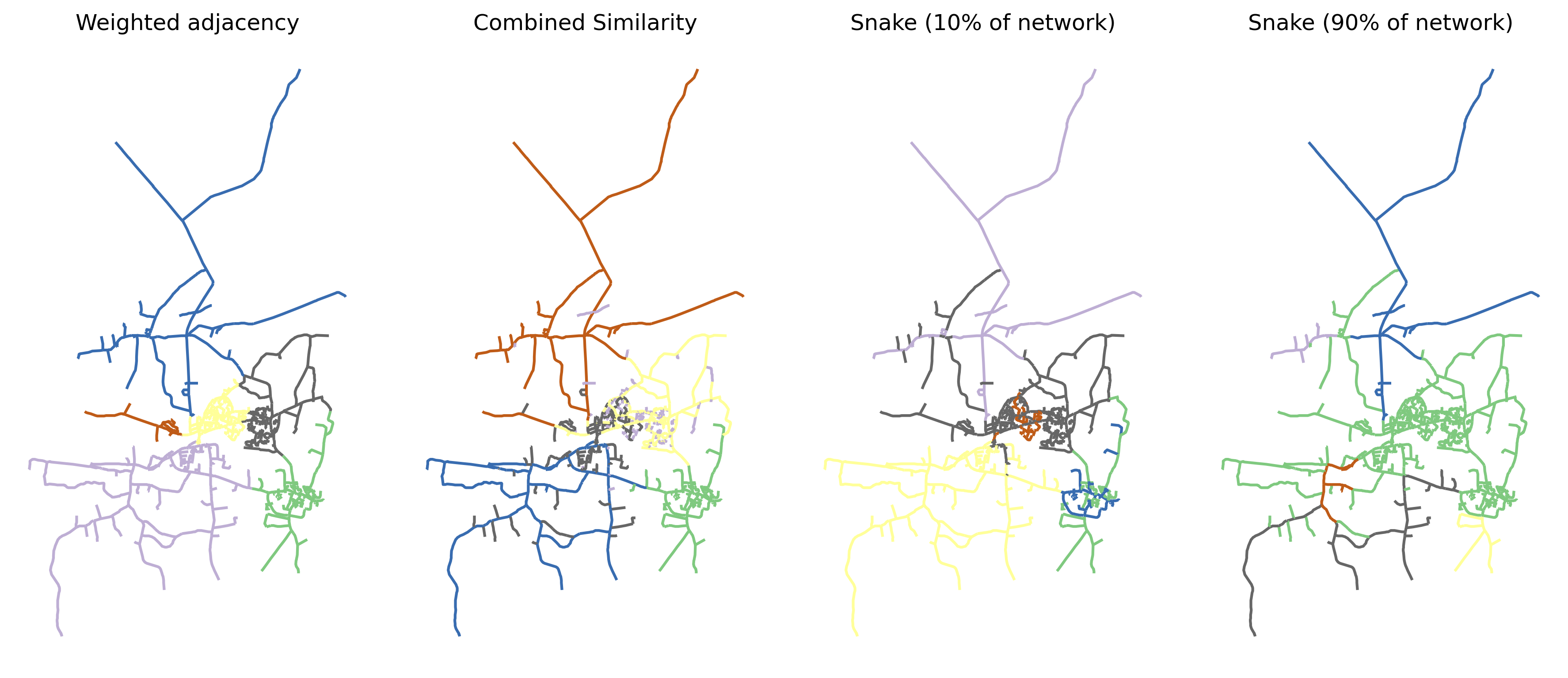}
    \caption{Results of spectral clustering into 6 groups using different inputs: weighted adjacency, combined similarity, and Snake-based similarity measures (with lengths equal to 10\% and 90\% of the network length).}
    \label{representation_clustering}
\end{figure}

The aforementioned representations (weighted adjacency, combined similarity, and Snake similarity) can also be compared based on their results when used as inputs to the same clustering method. We consider spectral clustering, which belongs to the same family as Symmetric Non-negative Matrix Factorization used to cluster Snakes \cite{saeedmanesh2016clustering}, with a fixed number of clusters set to 6 for illustration purposes. By only changing the representation, the clustering results shown in Figure~\ref{representation_clustering} vary substantially.

Spectral clustering on the weighted adjacency matrix yields very spatially compact clusters that are not strongly driven by attribute information. In contrast, spectral clustering on the combined similarity produces a more hierarchical structure, where, for instance, the central–eastern area consists of major axes (in light yellow) and smaller streets (in grey), patterns that can also be observed in other parts of the city. Using Snake similarity with a length equal to 10\% of the network length leads to similar conclusions, with some geographical areas exhibiting mixed clusters. Larger Snakes (90\% of the network length) result in a large cluster covering the central area, together with smaller clusters in suburban areas.

The choice of representation is therefore crucial, as it can lead to substantially different results depending on the strategy and the parameters used. The selection of a representation should be guided by the objective of the partitioning task: community detection may rely more heavily on attribute information, whereas road network partitioning may impose additional constraints such as connectivity. It is also important to note that this choice constitutes the first step toward the partitioning procedure, which, depending on the method, may in turn allow certain representation parameters to be tuned by optimizing an objective function.

Using pairwise distances (or their similarity equivalents) provides a flexible and versatile framework for handling a wide range of attribute types. While this section presented results where topological and attribute distances are combined into a unified matrix, the following section introduces the main contribution of this work: a method that jointly optimizes structural and attribute objectives without relying on a predefined combined representation, thereby offering greater flexibility in its application.

%% file: sections/4_proposed_methods.tex
\section{Optimal Transport-based Partitioning of Attributed Transportation Graphs}\label{srfgw}

The problem of partitioning an attributed graph lies in defining a partition that jointly account for both the graph structure and the node attributes. Existing methods mainly rely on an early-fusion strategy, following the categorization proposed in the comprehensive survey by \cite{chunaev2020community}. In this approach, structural and attribute information are combined into a common representation prior to partitioning (for instance by incorporating attribute information into the graph through edge weights), even when they exhibit different patterns.

The methodology proposed in this article differs to this approach as it belongs to the class of simultaneous fusion methods, which integrate structural and attribute information directly within the community detection process, for instance by modifying the objective function of the underlying algorithm. The main objective is to propose a method that enables a more precise control of the balance between attributes and topology, thereby providing greater versatility across a wide range of applications. In our approach, this is achieved through optimal transport (OT), whose objective is to map the input network onto a smaller target graph, thereby inducing a partition of the nodes into distinct communities.

\subsection{Graph Partitioning with semi-relaxed Fused Gromov–Wasserstein Discrepancy}

\subsubsection{Gromov--Wasserstein: Matching Graphs without Node Correspondence}

Matching two graphs amounts to comparing their representations, typically matrices, that do not share the same ground space, since they don't have the same number of nodes and because there is no natural notion of point-to-point distance between nodes belonging to different graphs. To address this issue, \cite{peyre2016gromov} propose a framework based on the Gromov--Wasserstein (GW) discrepancy \cite{memoli2011gromov}, which enables comparison of such structured objects. This formulation can be applied to graphs by representing a graph $G$ as a pair $(\mathbf{R}, \boldsymbol{\mu})$, where $\mathbf{R} \in \mathbb{R}^{N \times N}$ encodes the relationships between nodes (here, shortest-path distances between road segments), and $\boldsymbol{\mu} \in \Sigma_N$, where $\Sigma_N$ denotes the probability simplex. The distribution $\boldsymbol{\mu}$ assigns a weight to each node, reflecting its relative importance. In road networks, where all road segments are considered equally important, $\boldsymbol{\mu}$ is naturally chosen as the uniform distribution. In other applications, however, it may encode prior information such as node importance or usage frequency.

The GW discrepancy have broad applicability and could be utilized for various tasks. For graph partitioning, the main idea is to match a source graph $G^{(s)}$, represented by $(\mathbf{R}^{(s)}, \boldsymbol{\mu}^{(s)})$, with a much smaller target graph $G^{(t)}$ of $k$ nodes, each corresponding to a cluster. The target graph is represented by $(\mathbf{R}^{(t)}, \boldsymbol{\mu}^{(t)})$. The resulting optimal transport plan provides a soft assignment of each node to the super-nodes of the target graph, thereby defining the final partitions. 

\begin{equation*}
    GW_2(\textbf{R}^{(s)}, \boldsymbol{\mu}^{(s)}, \textbf{R}^{(t)}, \boldsymbol{\mu}^{(t)}) = \min_{\substack{\mathbf{T}\mathbf{1}_k = \boldsymbol{\mu}^{(s)} \\ \mathbf{T}^T \mathbf{1}_N = \boldsymbol{\mu}^{(t)}}} \sum_{i,j,l,m} |R^{(s)}_{ij} - R^{(t)}_{lm}|^2 T_{il}T_{jm}
\end{equation*}
with $\mathbf{1}_N$ is a column vector of size $N$ where all entries are equal to $1$ and $\mathbf{T} \in \mathbb{R}^{N \times k}_+$ the optimal transport plan, representing the probabilistic matching of nodes.

$\boldsymbol{\mu}^{(t)}$ require prior knowledge of the relative importance of the classes, information that is often unknown and unconstrained in clustering scenarios. To address this, \cite{vincent2021semi} propose relaxing the second marginal and introduce a semi-relaxed Gromov--Wasserstein (srGW) discrepancy. However, GW and srGW metrics focuses solely on structure and are therefore not directly suitable for attributed graphs. \cite{titouan2019optimal} proposed a new distance for structured data such as attributed graphs, called Fused Gromov--Wasserstein (FGW), which incorporates both topological and feature information, and which was also extended in a semi-relaxed setting (srFGW):

\begin{equation*}
srFGW_{2,\alpha}(\mathbf{R}^{(s)}, \boldsymbol{\mu}^{(s)}, \mathbf{R}^{(t)}, M_{AB}) = \min_{\substack{\mathbf{T}\mathbf{1}_k = \boldsymbol{\mu}^{(s)}}} 
\sum_{i,j,l,m} \Bigl(
    (1 - \alpha)  d_A(v_i^{(s)}, v_l^{(t)})^2
    + \alpha \, |R^{(s)}_{ij} - R^{(t)}_{lm}|^2
\Bigr) T_{il}T_{jm}
\end{equation*}

where $M_{AB} = (d_A(v_i^{(s)}, v_l^{(t)}))_{il}$ denotes the $N \times k$ distance matrix between $A$ and $B$, corresponding to the set of attributes of the source and target graphs, respectively. Therefore, the FGW distance seeks the optimal coupling that minimizes a linear combination of the cost of transporting node features from one graph to another and the cost of aligning pairs of nodes according to their respective graph structures.

\subsubsection{srFGW for Attributed Graph Partitioning}

Following \cite{gavra2025optimal}, we propose to use srFGW for partitioning attributed graphs, in particular through an appropriate specification of the target structure. This approach has the advantage of jointly accounting for both structure and attributes within the optimization process. 

As specified above, we consider $\mathbf{R}^{(s)} = \mathbf{D}_S$, the shortest-path distance matrix of the source graph, and $\boldsymbol{\mu}^{(s)}$ the uniform distribution. The target representation is defined as $\mathbf{R}^{(t)} = \delta^{(t)} (\mathbf{1}_{k \times k} - \mathbf{I}_k)$, where $\delta^{(t)}$ denotes the chosen distance between target nodes. In this formulation, $\mathbf{R}^{(t)}$ is a distance matrix with zero intra-group distances and strictly positive inter-group distances. As shown in \cite{gavra2025optimal}, using the average distance of the source graph leads to better results in most cases, and we therefore set $\delta^{(t)} = \mathrm{mean}(\mathbf{D}_S)$ for the remainder of this work.

$M_{AB}$ denotes the $N \times k$ distance matrix between $A$, the set of attributes of the source graph, and $B$ the set of attributes of the target graph. In the case of partitioning, however, the attributes of the target graph are, by definition, unknown. In this setting, the optimization problem could be reformulated as:

\begin{align}\label{srfgw_optimization}
srFGW_{2,\alpha}&(\mathbf{R}^{(s)}, \boldsymbol{\mu}^{(s)}, \mathbf{R}^{(t)}, A, d_A) = \min_{\substack{\mathbf{T}\mathbf{1}_k = \boldsymbol{\mu}^{(s)} \\ B \in \mathcal{X}^{k}}} 
\sum_{i,j,l,m} \Bigl(
    (1 - \alpha)  d_A(v_i^{(s)}, b_l)^2 
    + \alpha \, |R^{(s)}_{ij} - R^{(t)}_{lm}|^2
\Bigr) T_{il}T_{jm}
\end{align}

with $B = (b_1, \dots, b_k)$ designs the barycentric attributes of each classes. The barycenter $b_l$ for each class $C_l$ may be selected among the nodes of the graph, specifically $b_l \in \argmin_b \sum_{i} d_A(v_i, b)^q T_{il}$.

The partitioning algorithm, detailed in Algorithm~\ref{algo_srfgw}, alternates between solving the FGW problem for fixed barycenters and updating the barycenters according to the soft clustering induced by the resulting transport plan.

\begin{algorithm}
    \small
    \caption{semi-relaxed Fused Gromov--Wasserstein for node-attributed graph clustering}
    \begin{algorithmic}
    \STATE \textbf{Input:} An attributed weighted graph $G = (V, E, A)$ characterized by its structural distance matrix $\mathbf{D}_S^{(s)}$ and its probability distribution $\boldsymbol{\mu}^{(s)}$, the maximum number of desired subsets $k$, weighting parameter $\alpha$, an initial transportation plan $\mathbf{T}^0$ and a maximal number of iterations
    \STATE \textbf{Initialization:}
        \STATE Create the target matrix $\mathbf{D}_S^{(t)} = \mathrm{mean}(\mathbf{D}_S^{(s)})(\mathbf{1}_{k \times k} - \mathbf{I}_k)$
        \STATE $n = 0$
        \STATE $\mathbf{T}^n = \mathbf{T}^0$
    \REPEAT
        \STATE Compute $B^n = (b_1^n, \dots, b_k^n)$ the weighted barycentric attributes of each subset
        \STATE  Compute $\mathbf{M} = \left[ d_A(v_i, b_j) \right]_{\substack{1 \leq i \leq N \\ 1 \leq j \leq k}}$, the distance matrix between the node attributes and the barycentric attributes of each class.
        \STATE Get optimal transport plan $\mathbf{T}^{n+1}$ from $srFGW_{2,\alpha}(\mathbf{D}_S,\boldsymbol{\mu}^{(s)},\mathbf{D}_S^{(t)},\mathbf{M})$
        \STATE $k = \#\{ l \in \{1, \dots, k\} \mid C_l \neq \emptyset \}$
        \STATE $n = n+1$
        \UNTIL{$T^n = T^{n-1}$ \textbf{ or } $n >$ \textit{maximal number of iterations}}
        \STATE (optional) Get subsets from the last transportation coupling:\\ $C_l^n = \left\{ i \;\middle|\;
        \forall m \in \{1,\dots,k\},\;
        \begin{cases}
       \mathbf{T}^n_{il} > \mathbf{T}^n_{im} \\
        \text{or } \left( \mathbf{T}^n_{il} = \mathbf{T}^n_{im} \text{ and } l \le m \right)
        \end{cases}
        \right\}$
    \STATE (optional) Update the barycentric attributes according to the last hard clustering step
    \STATE \textbf{return} Subsets $\{C_1, \dots, C_k\}$ where some $C_l$ may be empty, and their corresponding barycentric attributes.
    \end{algorithmic}
    \label{algo_srfgw}
\end{algorithm}

\subsubsection{Choice of Hyperparameters}

\paragraph{Initial Transportation Plan $\mathbf{T}^0$}

\cite{vincent2021semi} discuss the sensitivity of GW solvers to initialization, particularly in settings where the target structure provides limited information, such as graph partitioning problems. In such cases, the optimization procedure may become trapped in local optima, highlighting the importance of a carefully chosen initial transportation plan $\mathbf{T}^0$. A common strategy consists in initializing the solver using the output of an alternative clustering or partitioning algorithm. This approach is adopted by initializing the transport plan using $k$-means clustering results, as recommended by default in the POT library implementation \cite{flamary2021pot} used in our implementation. To reduce the sensitivity to random initialization, the algorithm may be run multiple times with different random seeds, retaining the solution with the lowest objective value.

\paragraph{Maximum Number of Classes $k$ and Weighting Parameter $\alpha$}

Both the maximum number of classes $k$ and the weighting parameter $\alpha$ can be selected either through domain-specific rules, via optimization of one or several criteria, or through a combination of both approaches. For instance, in road network partitioning, the number of classes is typically constrained to a reasonable range rather than chosen freely, and can subsequently be refined by minimizing an appropriate objective function. Similarly, the balance between structure and attributes may be fixed a priori or tuned based on a connectivity-related metric or other task-dependent criteria.

For attributed transportation graphs, two types of criteria may be considered, since the objective is to obtain a partition that simultaneously satisfies both structural and attribute-based clustering objectives. Combining such criteria is common in this context. For instance, in road network partitioning, \cite{alayasreih2025cross} propose jointly using connectivity $C_{\mathrm{conn}}$ and normalized total variance $TV_n$ to evaluate clustering quality.

Let $C = \{C_1, \dots, C_k\}$ denote the set of clusters. Connectivity is defined at the level of the whole partition by aggregating cluster-wise connectivity scores weighted by cluster size (or length):

\begin{equation*}
    C_{\mathrm{conn}} = \frac{1}{l_G} \sum_{C_i \in C} l_{\mathrm{cc}(C_i)}
\end{equation*}

where $l_{\mathrm{cc}}(C_i)$ denotes the length of the largest connected component within cluster $C_i$ and $l_G$ is the total length of the graph.

The normalized total variance is defined as:

\begin{equation*}
    TV_n = \frac{\sum_{C_i \in C} l_{C_i} \cdot \mathrm{var}(C_i)}{l_G \cdot \mathrm{var}(G)}
\end{equation*}

where $\mathrm{var}(C_i)$ denotes the variance of the attributes within cluster $C_i$, and $\mathrm{var}(G)$ is the variance of the attributes over the whole graph. $TV_n$ can be interpreted as a normalized version of classical clustering inertia, incorporating length-based weighting.

For other applications, such as community detection, where the network encodes relational or usage-based interactions, classical criteria can also be employed like. In particular, modularity \cite{newman2006modularity} can be used to assess the structural quality of a partition, while silhouette can be used to evaluate attribute homogeneity, as illustrated in Section~\ref{BSS}.

Note that in road network example, structural criteria is maximized, while attribute-based criteria (total normalized variance) are minimized. Smaller values of $k$ or larger values of $\alpha$ lead to stronger emphasis on structural criteria and higher attribute variance, and conversely. The following section illustrates the selection of these two hyperparameters using our running example.

\subsection{Comparison with Existing Methods}

The proposed method is compared with two state-of-the-art approaches presented in Section~\ref{method_state_of_art}, representing two important families of methods: the Louvain algorithm, applied to a weighted adjacency matrix whose edge weights incorporate both structural and attribute-based distances, and spectral clustering based on the Snake similarity. 

\subsubsection{Methods hyperparameters}

These methods differ in how the number of clusters is controlled. It can be specified directly, as in spectral clustering, constrained through a maximum number of clusters, as in srFGW, or indirectly influenced by another parameter, as in the Louvain algorithm through the resolution parameter. In Louvain, increasing the resolution generally leads to a larger number of smaller communities, whereas decreasing it favors fewer, larger communities. Similarly, the balance between structure and attributes may be specified directly through $\alpha$, the weighting parameter used in srFGW, or implicitly through the construction of the input representation, such as in Louvain where edge weights encode both structural and attribute-based similarities, or in spectral clustering where this balance is controlled indirectly via the Snake length. These differences are summarized in Table~\ref{methodes_hyperparameters}.

Unlike the proposed OT-based methods, the Louvain algorithm does not require an explicit initialization. However, due to its stochastic node ordering, it may produce slightly different partitions across runs. Spectral clustering relies on a final $k$-means step and therefore also requires initialization. To ensure a fair comparison, all methods involving a stochastic optimization or initialization were repeated 100 times, and the solution maximizing their respective objective function was retained. 

\begin{table}
\centering
\begin{tabular}{@{}>{\arraybackslash}m{3.8cm}>{\centering\arraybackslash}m{4.5cm}
                >{\centering\arraybackslash}m{4.5cm}
                >{\centering\arraybackslash}m{4.5cm}@{}}
\hline
 & Louvain & Spectral clustering & srFGW \\
\hline
Graph representation &
Weighted adjacency incorporating both structure and attributes &
Snake similarity &
$\mathbf{D}_S$ and $\mathbf{D}_A$ structural and attribute distance matrices \\
\hline
Number of clusters &
Indirectly controlled by the resolution parameter &
Specified &
Maximum number of clusters specified \\
\hline
Structure-attribute balance &
$\alpha$ (through weighted adjacency) &
Snake length &
$\alpha$ \\
\hline
\end{tabular}
\caption{Summary of input design choices and hyperparameter settings for each method.}
\label{methodes_hyperparameters}
\end{table}

\subsubsection{Results}

The three methods are applied to our illustrative case study. Since this application is inspired by a road network partitioning problem, only a small number of clusters is considered, namely $k \in [4, 8]$. For the Louvain algorithm, the resolution parameter is adjusted to produce a comparable range of cluster numbers, using $\mathrm{resolution} \in [0.02, 0.03, 0.04, 0.08, 0.10]$. A wide range of structure–attribute balances is considered, namely $\alpha \in [0.1, 0.9]$, where small values of $\alpha$ correspond to a higher weight on attributes and, conversely, large values emphasize structural information. For the Snake length and to ensure comparability, $1 - \mathrm{Snake}_{\mathrm{prop}} \in [0.1, 0.9]$ is considered, where $\mathrm{Snake}_{\mathrm{prop}}$ denotes the proportion of the network covered by each Snake. In this setting, small values of $\mathrm{Snake}_{\mathrm{prop}}$ correspond to a more local representation and therefore place greater emphasis on structural information. 

\paragraph{Impact of structure-attribute balance}

The balance between structural and attribute information, controlled through $\alpha$ or through the Snake length, has a different effect depending on the method. In the Louvain algorithm, $\alpha$ only modifies the weights of adjacent edges and therefore has a local impact on the graph topology. In contrast, srFGW operates on the full pairwise distance matrix, allowing $\alpha$ to influence relationships between all pairs of streets rather than only neighboring ones, resulting in a more global effect. For spectral clustering based on Snake similarity, the influence of the Snake length is less direct. Increasing the Snake length progressively captures larger-scale structural patterns, as illustrated previously, but the representation remains inherently local since each building step depends only on the neighborhood of the current Snake.

\begin{figure}
    \centering
    \includegraphics[width=\textwidth]{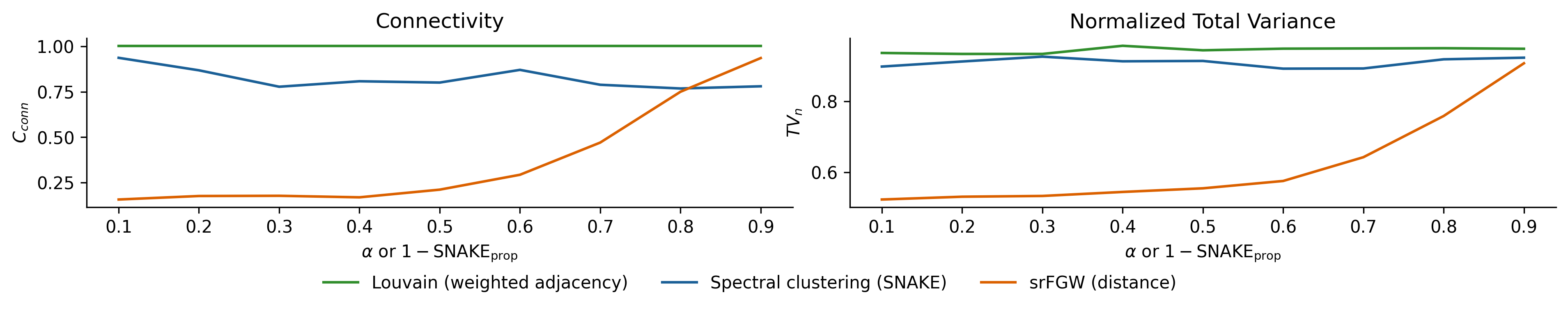}
    \caption{Evolution of connectivity and normalized total variance with respect to $\alpha$ and $1 - \mathrm{Snake}_{\mathrm{prop}}$ for $k=6$ and $\mathrm{resolution} = 0.04$}
    \label{alpha_impact}
\end{figure}

Figure~\ref{alpha_impact} illustrates this behavior for $k=6$, although similar trends are observed for both larger and smaller numbers of clusters. Overall, Louvain exhibits perfect connectivity results, as expected given that its objective function explicitly maximizes modularity. Increasing $\alpha$ or $1 - \mathrm{Snake}_{\mathrm{prop}}$ does not significantly improve connectivity for spectral clustering with Snake, which even shows a slight decrease. In contrast, srFGW is strongly affected by this parameter: while small values of $\alpha$ lead to poor connectivity, higher values allow the method to achieve results comparable to or better than spectral clustering with Snake.

For normalized total variance, srFGW achieves by far the best performance. Similarly to connectivity, Louvain and spectral clustering with Snake do not appear to be significantly affected by the evolution of $\alpha$ or $1 - \mathrm{Snake}_{\mathrm{prop}}$, which mainly influences local interactions. In contrast, srFGW is again strongly impacted, exhibiting an increase in variance when the balance shifts toward structural information. Overall, Figure~\ref{alpha_impact} illustrates that the fusion-based strategy in srFGW enables a more precise control of the balance between structure and attributes.

\paragraph{Impact of number of clusters} Theoretically, the number of clusters would not have a direct impact on the connectivity term, but it may play a more significant role in the normalized total variance. Indeed, increasing the number of clusters is expected to produce smaller and more homogeneous groups, thereby reducing attribute variance within clusters. 

\begin{figure}
    \centering
    \includegraphics[width=\textwidth]{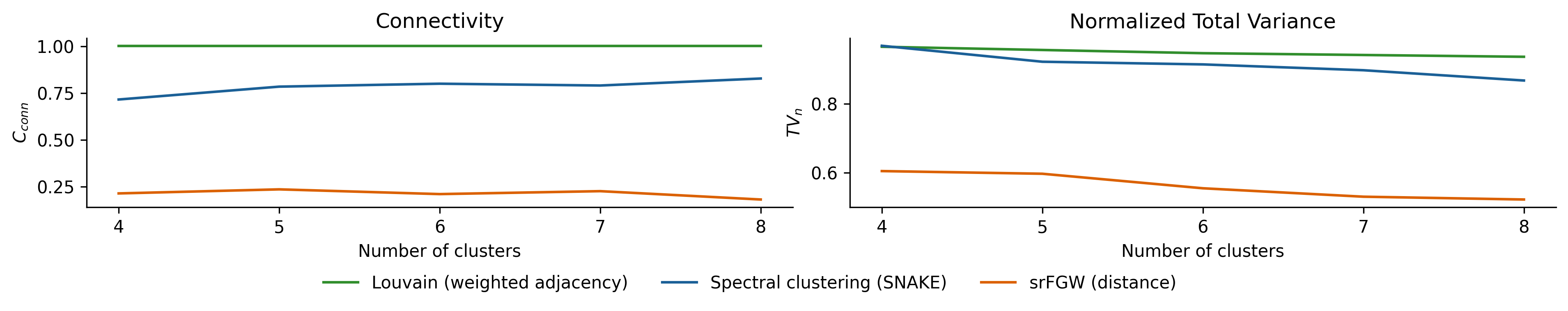}
    \caption{Evolution of connectivity and normalized total variance with respect to number of clusters for $\alpha=0.5$ and $1 - \mathrm{Snake}_{\mathrm{prop}}=0.5$}
    \label{k_impact}
\end{figure}

In Figure~\ref{k_impact}, the number of clusters does not play a significant role in connectivity across all methods, but has a decreasing impact on variance for both spectral clustering and srFGW. Overall, this parameter does not seem to have a major influence on performance in this setting.

\paragraph{Balancing Criteria}

The previous analyses highlight the trade-offs between the different methods. While Louvain performs well in terms of connectivity, it is less suitable when attribute homogeneity is a primary objective. Conversely, srFGW provides a substantial improvement in terms of normalized total variance, but exhibits weaker connectivity when small values of $\alpha$ place greater emphasis on attributes. Among the methods considered, srFGW is the most sensitive to the choice of $\alpha$ and, to a lesser extent, to the number of clusters.

To identify the most appropriate combination of hyperparameters, the criteria may be analyzed jointly or aggregated into a single objective, depending on the goals of the application. In the present case, a combined criterion is defined as the arithmetic mean of the connectivity and the complement of the normalized total variance ($1 - TV_n$) are first independently min--max normalized to the $[0,1]$ interval before being averaged with equal weights.

\begin{figure}
    \centering
    \includegraphics[width=\textwidth]{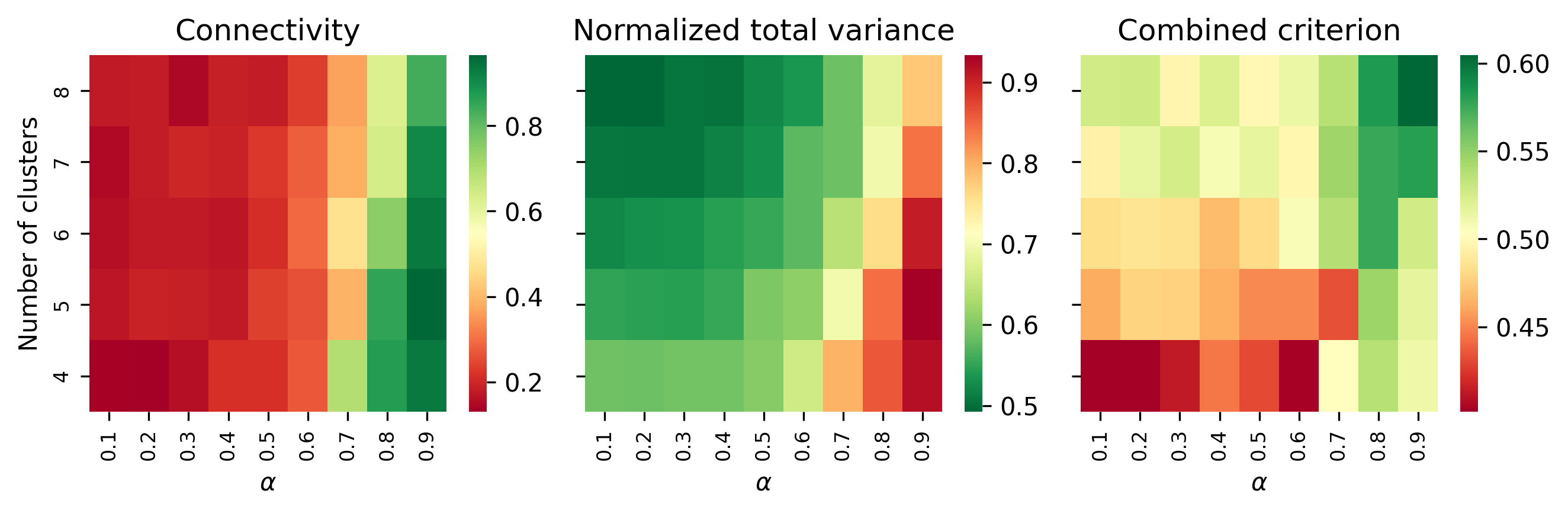}
    \caption{Heatmap of connectivity, normalized total variance and combined criterion with respect to number of cluster and $\alpha$.}
    \label{heatmap_criteria}
\end{figure}

The resulting criterion is illustrated in Figure~\ref{heatmap_criteria}. The same trends as in the previous analyses can be observed, including the influence of $\alpha$ on connectivity. The combined criterion further indicates that the best srFGW partitions are obtained for larger values of both the number of clusters and $\alpha$. However, using a large number of clusters would make the interpretation unnecessarily complex in this illustrative setting. Therefore, the remainder of the analysis focuses on partitions with six clusters and $\alpha = 0.8$.

\paragraph{Illustration of results}

\begin{figure}
    \centering
    \includegraphics[width=\textwidth]{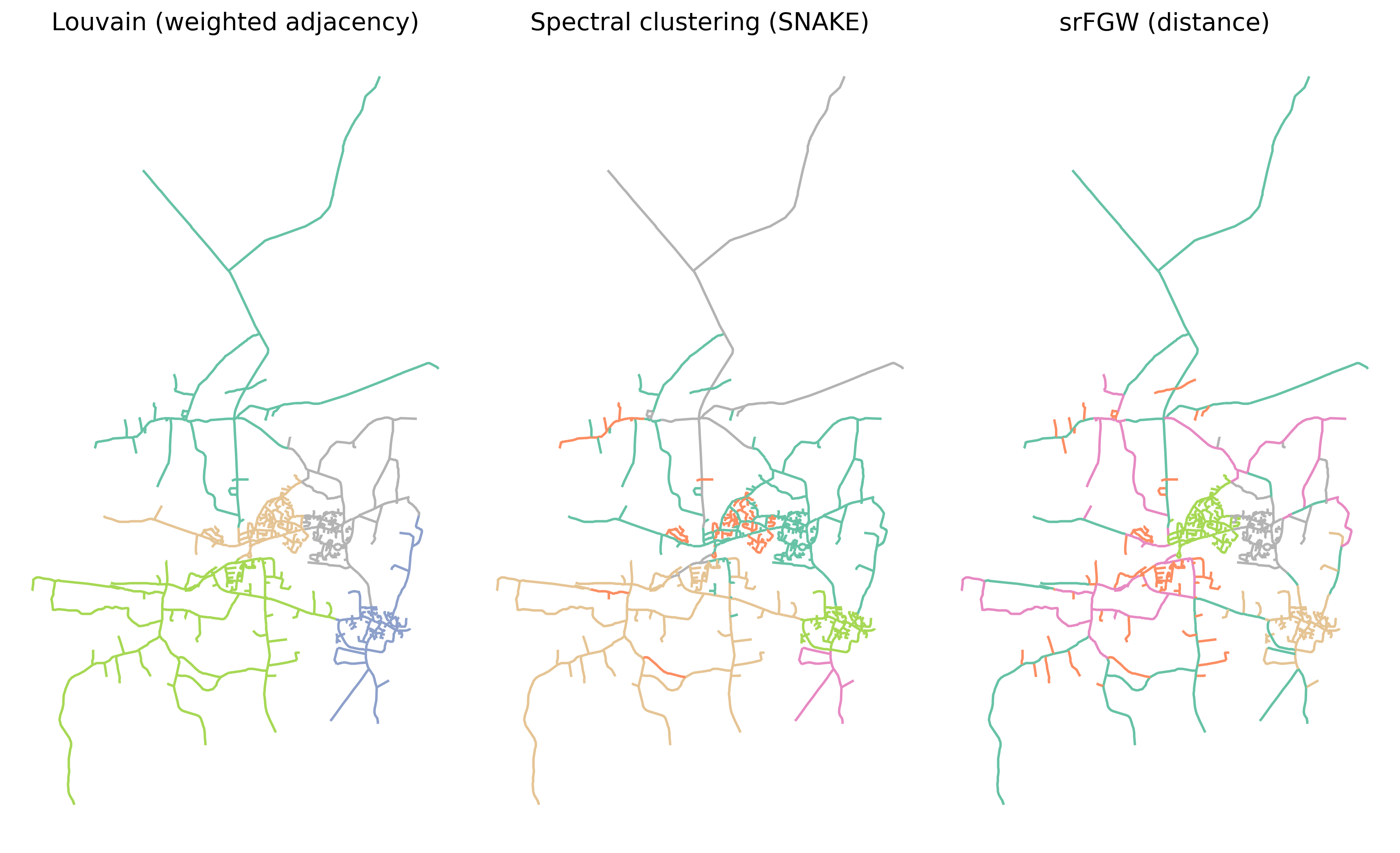}
    \caption{Partitions obtained by Louvain (resolution $= 0.04$, $\alpha = 0.8$), spectral clustering ($k = 6$, $\mathrm{Snake}_{\mathrm{prop}} = 0.2$), and srFGW ($k = 6$, $\alpha = 0.8$).}
    \label{methods_illustration}
\end{figure}

Figure~\ref{methods_illustration} illustrates the partitions obtained by the three methods for a comparable set of parameters. The Louvain algorithm produces connected clusters that are more driven by the topology of the road network than by the traffic attributes, as previously observed in the analysis. Spectral clustering with Snake and srFGW with distance matrices do not guarantee connectivity, and both yield spatially extensive clusters.

Spectral clustering appears to produce a partition separating the southern (beige) and northern (grey) areas, with a few smaller clusters, such as the pink one. The more interesting aspect concerns the segmentation of the city center, where some minor streets form a distinct (orange) cluster, while the remaining streets are grouped into a single (green) cluster.

The partition obtained with srFGW is more closely related to a typology of roads in the suburban area, with main exit roads (green), medium axes (pink), and dead ends (orange). In the city center, clusters appear to be more driven by the urban topology. This behavior reinforces previous results, showing that the method provides a balance between structure and attributes, and that this balance does not necessarily play the same role uniformly across the network.

%% file: sections/5_case_studies.tex
\section{Case Studies and Results}

The following section illustrates the proposed srFGW partitioning method through two case studies. The first case study investigates road network partitioning in the city center of Châteaubourg, introduced as the running example, by analyzing the traffic characteristics of the resulting clusters. The second applies the method to a bicycle-sharing usage network, illustrating its more general applicability to transportation networks and, more broadly, to community detection problems.

\subsection{Urban Road Network} \label{case:roadnetwork}

As highlighted in Section~\ref{running_example}, the city of Châteaubourg experiences recurrent congestion issues, mainly concentrated in its city center. Therefore, this illustrative case study focuses on this part of the city, excluding the suburban areas and the neighboring town to the southeast. While considering the whole city was relevant in the representation section due to the diversity of topological distances between rural and urban road segments, focusing on the city center provides a more representative road network partitioning case, although it is obviously smaller than typical urban networks. The resulting road network consists of 467 nodes and 1227 edges, corresponding to street segments and their intersections in the dual graph representation introduced previously. The nodes are associated with the same attributes as those described in Section~\ref{running_example}.

As in the previous analysis, several hyperparameters values were evaluated ($k \in [4,8]$ and $\alpha \in [0.1, 0.9]$, 100 repetitions) using the connectivity and normalized total variance criteria. The results presented below correspond to $\alpha = 0.7$ and $k = 8$, which provide the best compromise between the two criteria. In addition, selecting a relatively large number of clusters makes it possible to better illustrate the diversity of traffic patterns captured by the proposed method and to highlight the different traffic characteristics associated with each cluster.

\paragraph{Results}

\begin{figure}
    \centering
    \includegraphics[width=\textwidth]{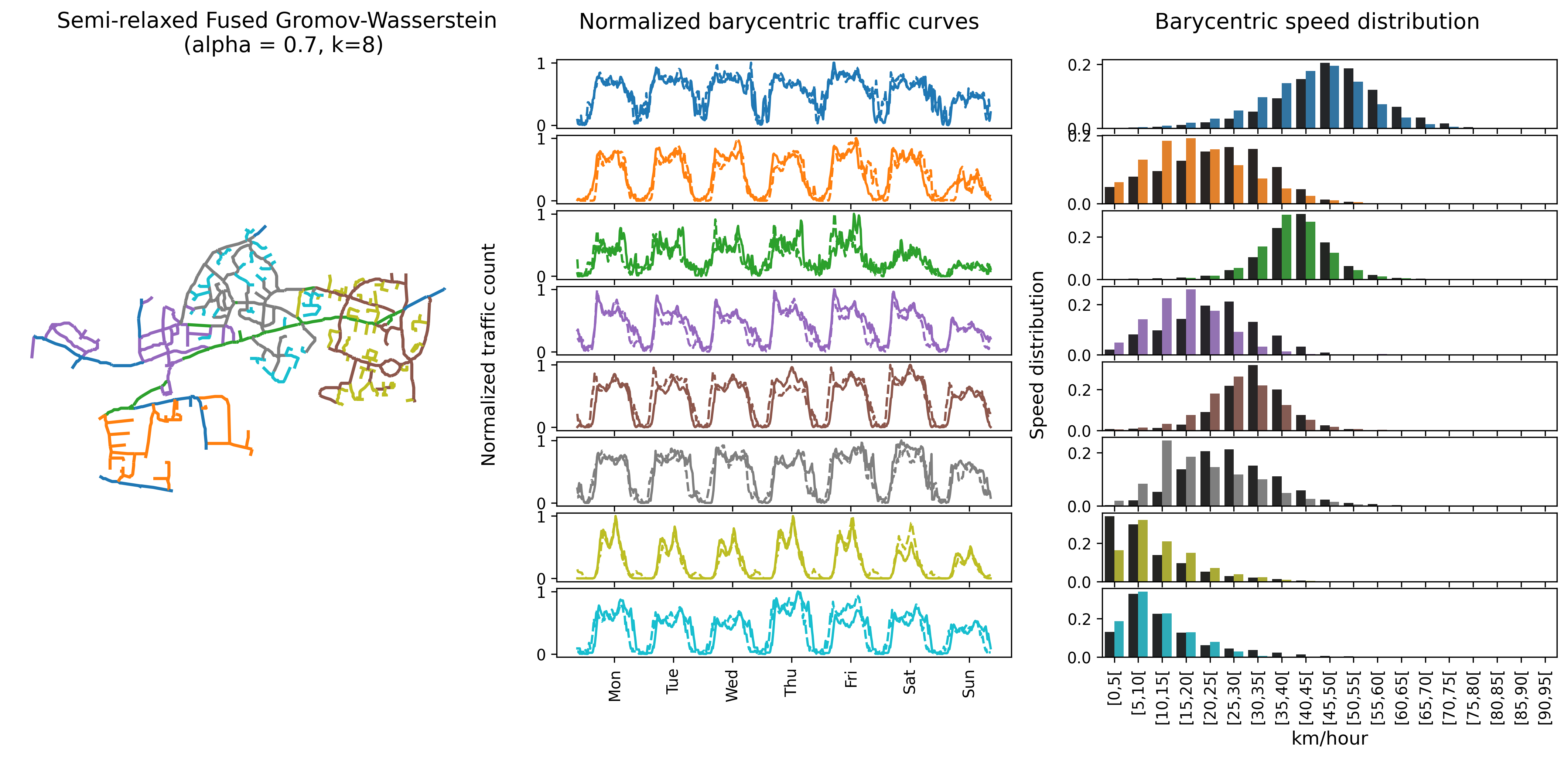}
    \caption{Clusters obtained by srFGW partitioning for $\alpha=0.7$ and $k=8$. }
    \label{ctb_partitiong_details}
\end{figure}

Figure~\ref{ctb_partitiong_details} presents the results as a clustered map of the city center together with the traffic attributes of one representative road segment for each cluster. The representative is chosen as the barycentric element, defined as the road segment minimizing the sum of distances to the other elements within its cluster.

The green cluster consists exclusively of the main road crossing the city, which experiences recurrent congestion. It is characterized by a pronounced decrease in traffic during weekends and strong commuter traffic throughout the rest of the week. Its speed distribution is among the highest, second only to the blue cluster. The latter mainly corresponds to the city's principal exit roads and, in contrast to the green cluster, exhibits a more uniform usage pattern throughout both the day and the week.

Orange and purple clusters concerns residential neighborhoods at different side of the city, characterised by similar speed distribution, which differ according to the direction of traffic and more pronounced morning peaks for purple cluster.

The remaining clusters form a classification of two areas of the city, distinguishing minor and medium-capacity roads. In the eastern part, the brown cluster separates medium-capacity roads, characterized by higher speed distributions, from the yellow cluster, which mainly corresponds to dead-end streets. Similarly, in the central area, the grey cluster represents medium-capacity roads, while the cyan clusters correspond to dead-end or residential streets. It can also be noted that the speed distributions of the grey and brown clusters differ, likely due to the presence of school-related activity in the grey area, which may contribute to traffic slowdowns.

It is worth noting that, despite all attributes being normalized and no explicit traffic volume information being provided, the resulting classification successfully reproduces a distinction between major, medium-capacity, and minor-capacity roads. The average peak-hour capacity is provided in Table~\ref{average_traffic} for reference.

\begin{table}
\centering
\begin{tabular}{cccccccc}
\hline
Green & Orange & Blue & Purple & Brown & Grey & Cyan & Yellow \\
\hline
429 & 113 & 111 & 36 & 35 & 23 & 6 & 5 \\
\hline
\end{tabular}
\caption{Average traffic volume for the streets in the corresponding cluster on a weekday between 4 p.m. and 6 p.m.}
\label{average_traffic}
\end{table}

Although the clusters do not guarantee connectivity, they exhibit strong spatial coherence within each cluster and effectively capture different functional areas of the city. Overall, these results illustrate the ability of srFGW to jointly consider both network topology and traffic attributes, regardless of their form (traffic curves, histograms, scalar values), and to produce informative partitions of the road network with limited parameter tuning.

\subsection{Bicycle Sharing System}\label{BSS}

To further illustrate the proposed method on real-world data and demonstrate its versatility, this second application considers another type of network, shaped by usage patterns rather than by geographical structure. This illustration focuses on the London bicycle-sharing system, namely the London Cycle Hire scheme (LCH), launched in 2010. The system has been extensively studied, as recently reviewed by \cite{zhang2024lessons}, mainly due to the large volume of open data made publicly available by its operator, Transport for London (TfL). 

These studies help to better understand travel patterns in order to adapt management strategies (bicycle relocation or service planning), identify bottlenecks and operational inefficiencies, and assess how events such as the COVID-19 pandemic affect cycling behavior at a broader scale. Some of this work has focused on community detection, such as \cite{munoz2018community}, which notably investigates travel flows between the resulting communities.

\subsubsection{Data Description}
 
This study relies on trip records of the TfL database for the whole year 2024, including the departure and arrival times and stations associated with each trip. Yearly data are aggregated at the station level to provide an overview of station usage. 

\paragraph{Stations Network}

\begin{figure}
    \centering
    \includegraphics[width=0.75\textwidth]{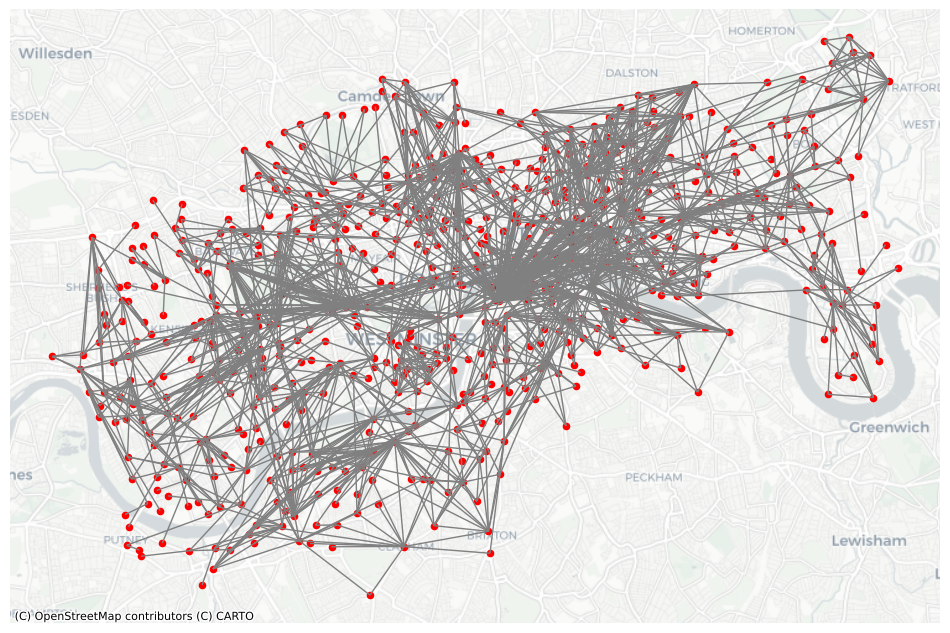}
    \caption{London Cycle Hire network, constructed from 2024 usage data.}
    \label{london_network}
\end{figure}

During 2024, around 800 stations were active according to the available data, although some were only sparsely used. Origin--destination trips are used to construct a network in which two stations are connected by an edge if at least 200 trips were recorded between them during the year, in either direction. This filtering step leaves some stations as isolated nodes or in connected components disconnected from the main network. As these stations already form distinct groups, they are excluded from the remainder of the study. The present study focuses exclusively on the largest connected component, which consists of 656 stations linked by 2454 edges. Figure \ref{london_network} proposes an overview of this final network.

Edges between stations may be left unweighted. However, depending on the intended interpretation of the network, several edge length definitions can be considered. These include a measure of usage intensity (e.g., the inverse of the number of trips, bringing stations that are frequently connected by users closer together), the physical distance between stations (either Euclidean or shortest-path distance on the road network), or a proxy for physical distance such as the median travel time between two stations. Following \cite{munoz2018community}, the present study characterizes edges by the intensity of usage, defining the edge length as the inverse of the number of trips recorded between the corresponding pair of stations.

\paragraph{Stations Data Details}

\begin{figure}
    \centering
    \includegraphics[width=\textwidth]{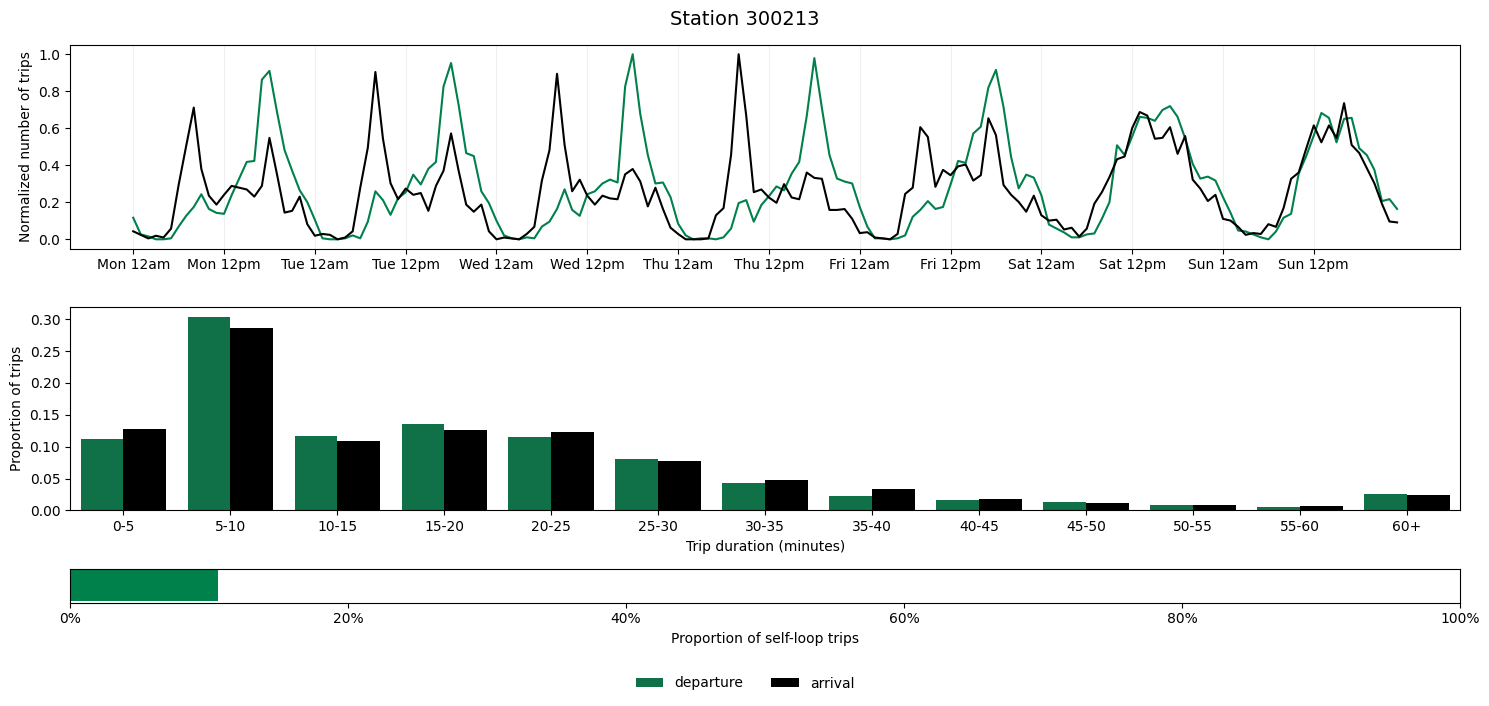}
    \caption{Illustration of the attributes associated with a station: normalized incoming and outgoing traffic, trip-duration distributions for trips to and from the station, and proportion of self-loop trips.}
    \label{london_attributes}
\end{figure}

First, the numbers of arrivals and departures by day of the week and hour are computed for each station, yielding average traffic profiles to and from the station. Second, trip duration are used to construct histograms of travel times to and from each station (using 5-minute bins). This type of histogram makes it possible to distinguish stations mainly associated with short trips from those characterized by a larger proportion of longer-duration usage. Finally, the proportion of self-loop trips is considered. Figure \ref{london_attributes} provides an illustration of the final aggregated data for a given station.

Other indicators could be considered depending on the objective of the study, such as the weekly evolution of occupancy rates, or the proportion of time during which a station is either completely empty or, conversely, full.

\paragraph{Stations Data Distances}

Unlike the road network partitioning case presented previously, departure and arrival information carry distinct meanings. Consequently, attribute distances are computed separately for departure and arrival profiles, ensuring that only quantities with the same semantic interpretation are compared, without resorting to the Hausdorff distance. For self-loop trips, represented by a single scalar value, the Euclidean distance is used. Overall, attributes are represented through five pairwise distance matrices: DTW distance matrices for arrival and departure curves, Wasserstein-1 distance matrices for arrival and departure duration histograms, and a Euclidean distance matrix for self-loop proportions. In this study, all attributes are assigned equal weights.

\paragraph{Evaluation Criteria}

Modularity and silhouette are classical indicators for community detection, the former evaluating structural coherence while the latter focuses on attribute consistency. Modularity is defined as:
\begin{equation*}
Q = \frac{1}{2m} \sum_{i,j} \left(A_{ij} - \frac{k_i k_j}{2m}\right)\mathbf{1}_{\{c_i = c_j\}}
\end{equation*}
where $A_{ij}$ is the adjacency matrix, $k_i$ is the degree of node $v_i$, $m$ is the total number of edges, and $c_i$ denotes the cluster assignment of node $v_i$. The term $\mathbf{1}_{\{c_i = c_j\}}$ is an indicator function equal to 1 when nodes $v_i$ and $v_j$ belong to the same cluster, and 0 otherwise. Larger values of modularity indicate a stronger community structure, corresponding to dense connections within clusters and sparse connections between them.

The clustering quality with respect to attributes is assessed using a silhouette-based criterion. For each node $v_i \in C_k$, the intra-cluster dissimilarity is defined as $a(v_i)=\frac{1}{|C_k|-1}\sum_{v_j \in C_k,\, j\neq i} d_A(v_i,v_j)$, while the inter-cluster dissimilarity is $b(v_i)=\min_{C_\ell \neq C_k} \frac{1}{|C_\ell|}\sum_{v_j \in C_\ell} d_A(v_i,v_j)$. The silhouette score is then $s(v_i)=\frac{b(v_i)-a(v_i)}{\max(a(v_i),b(v_i))}$, and the global criterion is obtained by averaging over all nodes as $\mathcal{S}=\frac{1}{n}\sum_i s(v_i)$, where $d_A(v_i,v_j)$ denotes the attribute distance between nodes $v_i$ and $v_j$. Larger silhouette values indicate that nodes are, on average, more similar to the members of their own cluster than to those of neighboring clusters, reflecting better attribute-based separation.

\begin{figure}
    \centering
    \includegraphics[width=\textwidth]{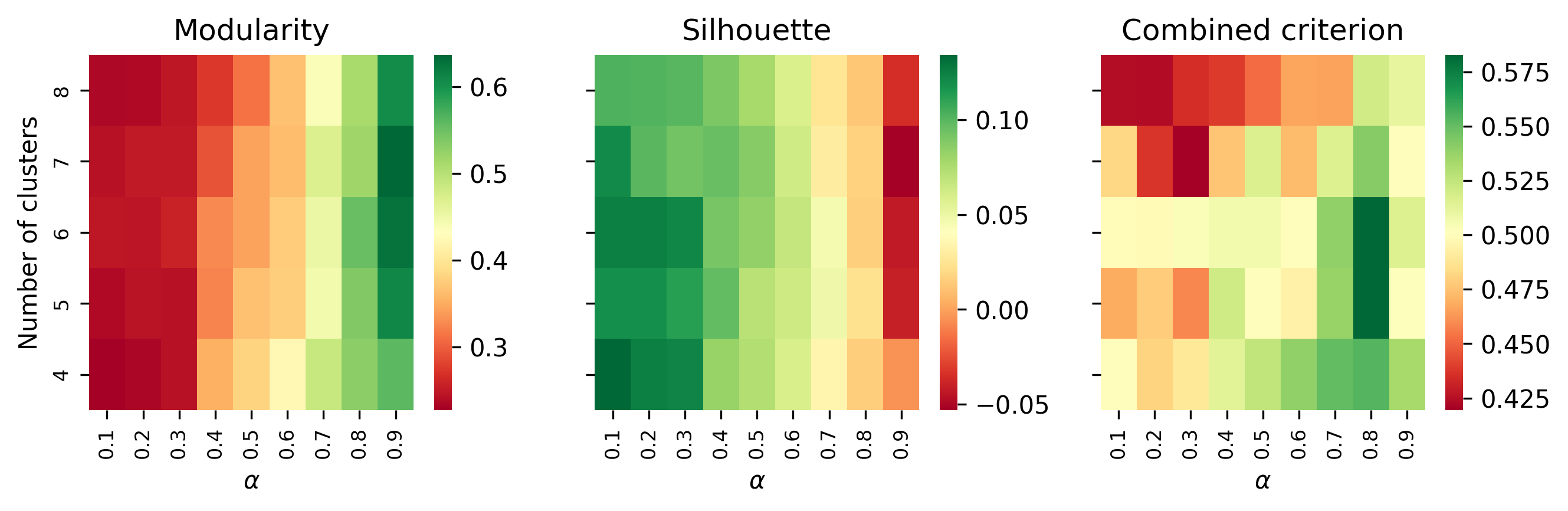}
    \caption{Heatmap of modularity, silhouette and combined criterion with respect to number of cluster and $\alpha$.}
    \label{london_heatmap_criteria}
\end{figure}

\begin{figure}
    \centering
    \includegraphics[width=\textwidth]{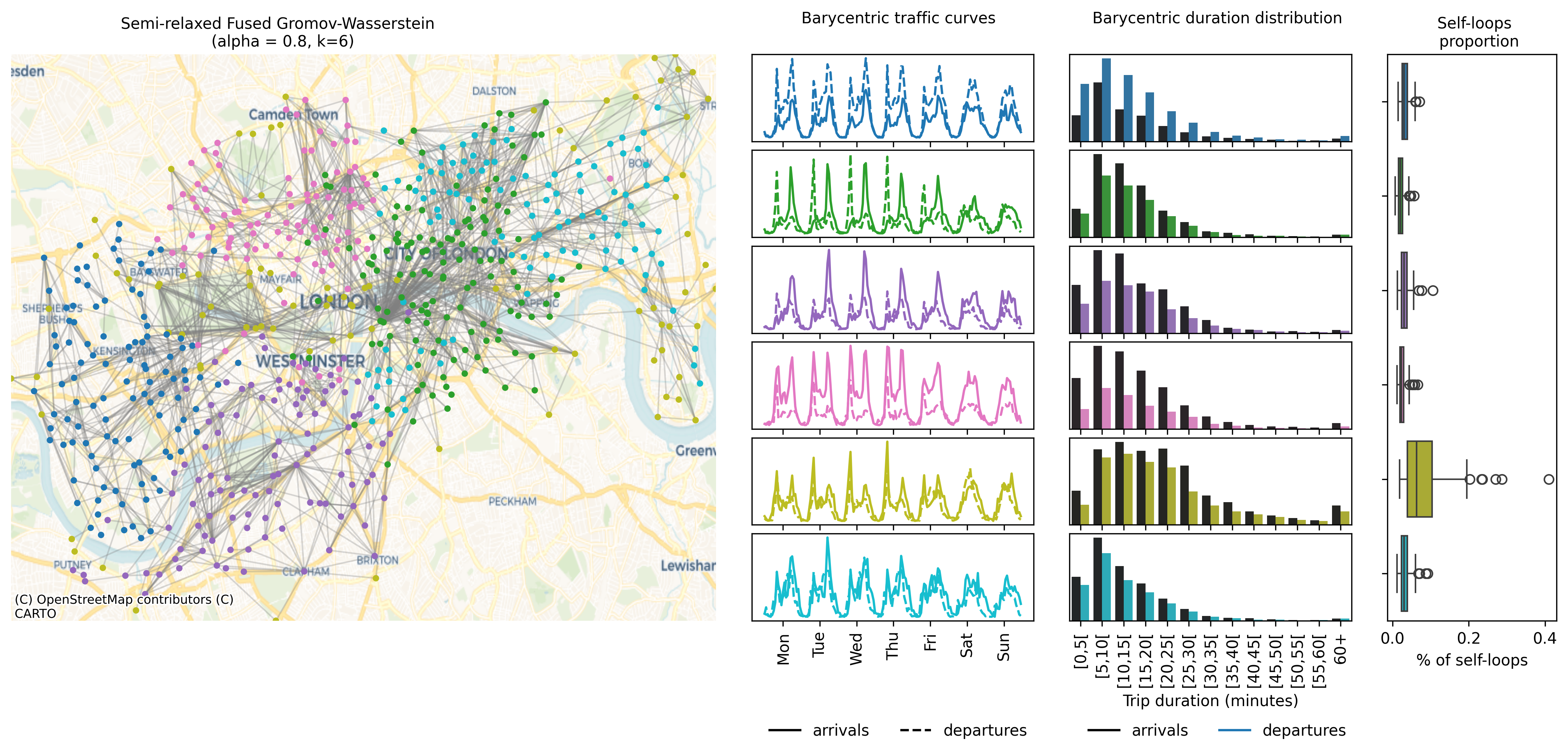}
    \caption{Clusters obtained by srFGW partitioning for $\alpha=0.8$ and $k=6$. }
    \label{london_partitiong_details}
\end{figure}

\paragraph{Results}
Following the approach adopted in the previous analysis, hyperparameter values ($k \in [4,8]$ and $\alpha \in [0.1,0.9]$) were evaluated over 100 repetitions using the modularity and silhouette criteria. Figure~\ref{london_heatmap_criteria} presents these criteria together with their equally weighted combined score. As in the previous case study, the number of clusters has only a limited impact on the results, except for very small or very large values of $k$. In contrast, the balance parameter $\alpha$ has a much stronger influence: small values lead to poor modularity but high attribute coherence, whereas large values exhibit the opposite behavior. The combined criterion identifies $\alpha = 0.8$ as the best compromise. For this value of $\alpha$, both $k=6$ and $k=7$ achieve similarly good results. The remainder of the analysis focuses on $k=6$, providing a sufficiently detailed partition while facilitating comparison with previous studies. In particular, \cite{munoz2018community} identified six communities using the Infomap algorithm, while \cite{zaltz2013structure} identified five to six communities using a modularity-based method. Both studies modeled LCH stations as a graph without incorporating attributes. The resulting partition is detailed in Figure~\ref{london_partitiong_details}.

LCH stations are partitioned into six clusters, which are mostly spatially coherent. However, some clusters include geographically distant stations, as they are characterized more strongly by their usage attributes than by their topological position in the network. In particular, the yellow cluster includes stations located at the periphery of the network, as well as stations close to major parks such as Hyde Park, Regent's Park, and Victoria Park. Previous studies based on these data have already highlighted specific usage patterns around Hyde Park, particularly during weekends \cite{zaltz2013structure, munoz2018community}. However, the grouping of peripheral stations and park-adjacent stations within the same cluster reveals a different type of organization, enabled by the optimal transport formulation. Indeed, this approach allows stations to be grouped according to their structural and attribute similarities rather than solely by their usage proximity. In this case, the clustering is also influenced by self-loop proportion indicators: the traffic profiles of the barycentric stations of this cluster are characterized by a large proportion of self-loop trips, higher activity during weekends, and longer trip durations. The other spatially dispersed cluster is the cyan cluster, which exhibits the smallest variations between weekdays and weekends. This cluster mainly corresponds to the remaining areas of Tower Hamlets, but also includes stations located in Islington and Lambeth, which are dense central districts characterized by diverse daily cycling uses.

Other clusters appear to be more geographically compact. In particular, the western part of London, which was identified as a distinct area in \cite{munoz2018community} and \cite{zaltz2013structure}, is divided in the proposed partition into two clusters: the blue cluster, composed of Kensington, Chelsea, Hammersmith and Fulham, and the purple cluster, corresponding to Wandsworth. Although this distinction was not observed in \cite{munoz2018community} and both clusters share similar residential characteristics, their cycling patterns suggest different functions within the bike-sharing network. The blue cluster is characterized by strong commuting-related patterns, with pronounced morning and evening peaks and a higher number of departures than arrivals, suggesting a predominantly residential origin area. Conversely, Wandsworth exhibits a stronger destination-oriented pattern, with marked evening arrival peaks. Moreover, this cluster is distinguished by differences in trip duration distributions between arrivals and departures, with shorter trips observed for inbound journeys. This suggests that Wandsworth may act as a local destination within the network, where users tend to make shorter final-leg trips when arriving in the area.

The pink cluster (mainly located to Camden) forms a cluster characterized by a dual mobility pattern. During weekdays, strong morning and evening arrival peaks suggest the influence of commuting and daily activities, with the area acting as an urban destination. In contrast, weekend arrivals are more evenly distributed throughout the day, with a marked increase around midday, reflecting leisure, cultural, and tourism-related activities. This temporal shift highlights the mixed function of this cluster, combining a regular activity center with a major recreational destination.

The green cluster mainly consists of stations located in Central London, including the City as well as major railway stations such as Waterloo, King’s Cross and St Pancras. This cluster is characterized by a high number of departures in the morning and arrivals in the evening, revealing a strong commuter-oriented pattern. It has the sharpest decrease in activity during weekends among all clusters, further supporting the interpretation of this area as a major transport hub, where cycling demand is closely linked to working hours.

\begin{figure}
    \centering
    \includegraphics[width=\textwidth]{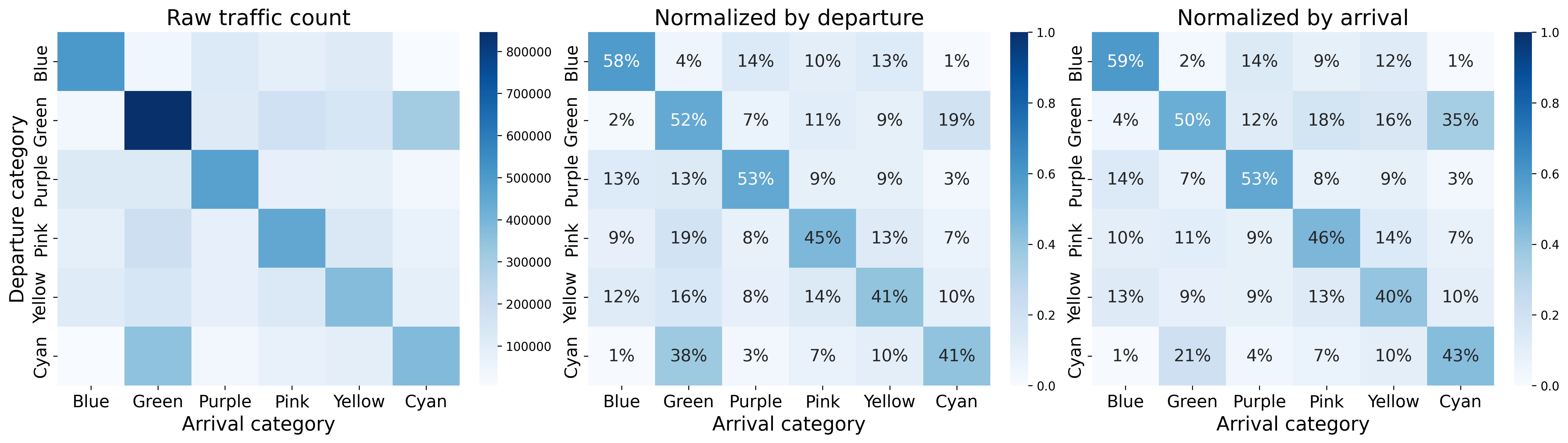}
    \caption{Inter-cluster traffic flow patterns and their normalized distributions.}
    \label{london_exchanges}
\end{figure}

Finally, Figure~\ref{london_exchanges} highlights the traffic patterns between the defined clusters. The proportion of intra-cluster trips is high, particularly for the blue, green, and purple clusters, but also across all clusters overall. This result is consistent with the high value of $\alpha$ selected, which favors partitions with strong intra-cluster connectivity. Some clusters exhibit limited exchanges with other groups, such as the exchanges between the blue and cyan clusters, whereas others share a large number of trips, notably the green and cyan clusters.

Overall, combining traffic attributes with the usage network structure reproduces some of the patterns observed in previous analyses of these data, while further refining the partition by subdividing existing areas or identifying clusters characterized by their arrival and departure profiles or trip durations. Moreover, incorporating traffic curves enables the consideration of the entire week within a single analysis, instead of separately analyzing weekdays and weekends.

\FloatBarrier
\section{Conclusion and Discussion}

This paper introduced a flexible framework for partitioning heterogeneous transportation networks by jointly considering network structure and transportation attributes. The proposed framework was not designed for a specific transportation application, but rather developed as a generic approach capable of integrating diverse sources of information. Its flexibility was then illustrated through several case studies with different objectives and network structures. Our approach relies on a distance-based representation of attributed graphs, allowing different types of transportation information to be incorporated, including attributes that are not restricted to scalar values. This representation provides a general way to characterize the complexity of transportation networks and to define dissimilarities between their components. 

A key contribution of the proposed framework is its ability to explicitly control the balance between structural and attribute information during partitioning. Unlike approaches relying on a predefined combination of these sources of information, the proposed methodology allows the emphasis given to network topology and operational characteristics to be adapted according to the application objective. This flexibility is particularly relevant in transportation systems, where meaningful partitions may depend on whether the objective is to preserve connectivity, identify homogeneous operational patterns, or reveal usage-based structures.

The use of optimal transport also provides additional interpretability through representative cluster barycenters learned during the optimization process. These barycenters summarize the characteristics of the obtained partitions and facilitate the analysis of how different structural and attribute preferences influence the resulting organization of the network. Moreover, the Gromov--Wasserstein formulation provides a relational perspective by characterizing each network component through its connections with the entire graph. As a result, components with similar structural roles, including those that are globally distant from most other components, can be assigned to the same partition. This enables the identification of specific structural configurations that may not be detected by approaches relying only on local similarity measures.

Several limitations should nevertheless be acknowledged. First, the proposed distance-based representation may become computationally demanding for very large transportation networks, particularly when complex attributes requiring costly distance computations are considered. Future research could investigate more scalable approximations, including faster alternatives for computing structural distances. Second, the current formulation does not explicitly enforce connectivity constraints within the resulting partitions. However, such constraints could be incorporated through additional regularization terms in the optimization objective when spatially connected regions are required. Finally, although the framework can integrate temporal or functional attributes, the current methodology does not explicitly address the dynamic evolution of transportation networks. Extending the approach toward dynamic or adaptive partitioning represents an important direction for future research.

Overall, this work provides a general framework for transportation network partitioning that accommodates heterogeneous information sources while offering explicit control over the trade-off between structure and attributes. By separating the representation of transportation information from the partitioning objective, the proposed approach can be adapted to a wide range of transportation applications.